\documentclass{aa}
\usepackage{graphicx}
\usepackage[varg]{txfonts}
\usepackage{natbib}
\newlength{\linwx}
\usepackage{color}
\begin{document}

\title{The structure of protoplanetary discs around evolving young stars}
\author{
Bertram Bitsch \inst{1},
Anders Johansen \inst{1},
Michiel Lambrechts \inst{1},
\and
Alessandro Morbidelli \inst{2}
}
\offprints{B. Bitsch,\\ \email{bert@astro.lu.se}}
\institute{
Lund Observatory, Department of Astronomy and Theoretical Physics, Lund University, 22100 Lund, Sweden
\and
University Nice-Sophia Antipolis, CNRS, Observatoire de la C\^{o}te d'Azur,Laboratoire LAGRANGE, CS 34229, 06304 NICE cedex 4, FRANCE
}
\abstract{
The formation of planets with gaseous envelopes takes place in protoplanetary accretion discs on time scales of several million years. Small dust particles stick to each other to form pebbles, pebbles concentrate in the turbulent flow to form planetesimals and planetary embryos and grow to planets, which undergo substantial radial migration. All these processes are influenced by the underlying structure of the protoplanetary disc, specifically the profiles of temperature, gas scale height, and density. The commonly used disc structure of the Minimum Mass Solar Nebula (MMSN) is a simple power law in all these quantities. However, protoplanetary disc models with both viscous and stellar heating show several bumps and dips in temperature, scale height, and density caused by transitions in opacity, which are missing in the MMSN model. These play an important role in the formation of planets, since they can act as sweet spots for forming planetesimals via the streaming instability and affect the direction and magnitude of type-I migration. We present 2D simulations of accretion discs that feature radiative cooling and viscous and stellar heating, and they are linked to the observed evolutionary stages of protoplanetary discs and their host stars. These models allow us to identify preferred planetesimal and planet formation regions in the protoplanetary disc as a function of the disc's metallicity, accretion rate, and lifetime. We derive simple fitting formulae that feature all structural characteristics of protoplanetary discs during the evolution of several Myr. These fits are straightforward for applying to modelling any growth stage of planets where detailed knowledge of the underlying disc structure is required.
}
\authorrunning{Bitsch et al.}\maketitle
%
%

\section{Introduction}
\label{sec:introduction}

Protoplanetary discs evolve over million year time scales during which the accretion rate $\dot{M}$ onto the central star drops, from typically $5\times 10^{-6}M_\odot$/yr around the youngest stars to $10^{-9}M_\odot$/yr after $3-10$ Myr \citep{1998ApJ...495..385H}. This reduction in $\dot{M}$ can be interpreted as a reduction of the gas surface density $\Sigma_G$. Ensembles of protoplanetary discs can be observed in star forming regions, like the Ophiuchus cluster \citep{2010ApJ...723.1241A}. These observations give constraints on the structure of protoplanetary discs (e.g. gradients in surface densities), although with large uncertainties. In the late stages of the discs, $\dot{M}$ becomes so small that the disc is blown away by photo evaporation on a very short time scale \citep{2013arXiv1311.1819A}. The central star also evolves on these time scales \citep{1998A&A...337..403B}. As the stellar luminosity changes, the amount of stellar heating received by the disc changes as well, which in turn affects the temperature and density structure.

During this evolution, planets form. Gas giant planets have to form, while the gas-rich protoplanetary disc is still present, and large parts of the growth of terrestrial planets and super Earths occur during the gaseous disc phase as well. Several important growth and formation stages rely on detailed knowledge of the protoplanetary disc structure:

\begin{itemize}
 \item Growth of dust particles to pebble sized objects \citep{2010A&A...513A..57Z, 2012A&A...539A.148B, 2013A&A...552A.137R}
 \item Movement of pebbles inside the disc due to gas drag \citep{1977Ap&SS..51..153W,2008A&A...487L...1B}
 \item Formation of planetesimals, e.g. via the streaming instability \citep{2005ApJ...620..459Y, 2007ApJ...662..627J}
 \item Turbulent stirring of planetesimals \citep{2009Icar..204..558M, 2009ApJ...707.1233Y, 2012ApJ...748...79Y}
 \item Formation of planetary cores from embryos and planetesimals \citep{2010AJ....139.1297L} or via pebble accretion \citep{2012A&A...544A..32L} and subsequent gas accretion onto them \citep{1996Icar..124...62P}
 \item Migration of planets and cores in the gas disc \citep{1997Icar..126..261W, 2006A&A...459L..17P, 2008A&A...487L...9K, 2009A&A...506..971K}.
 \item The late stages of the disc evolution, where migration of newly formed gas giants shape the inner solar system \citep{2009Icar..203..644R, 2011Natur.475..206W}
\end{itemize}

Even though these mechanisms happen on different length scales and time scales, they are all strongly dependent on the underlying disc structure (temperature $T$, gas surface density $\Sigma_G$, aspect ratio $H/r$), which makes the protoplanetary disc structure therefore a key parameter to understand the formation of planets and planetary cores. We now highlight some key processes, which will be discussed further in this paper. 

{\it Planetesimal formation} and the formation of planetary embryos can be aided by reducing the radial pressure gradient in the protoplanetary disc. Bumps and dips in the surface density of the disc, which is where the pressure gradient changes, can locally reduce the pressure support of the gas. A reduction in the pressure support of the gas causes a reduction of the headwind acting on the pebble and would reduce their inward motion \citep{1977MNRAS.180...57W, 2008A&A...487L...1B, 2012A&A...539A.148B}. Reduced pressure support also decreases the metallicity threshold for particle concentration through the streaming instability, significantly helping the formation of planetesimals \citep{2010ApJ...722L.220B}.

{\it Core accretion} is the process where planetary embryos grow to the size of the cores of the giant planets ($\sim$10 M$_{\rm Earth}$). The accretion of planetesimals by embryos is slower than disc life times \citep{1996Icar..124...62P, 2004astro-ph0406469, 2010AJ....139.1297L}, but growth time scales can be drastically reduced by considering the accretion of pebbles \citep{2010MNRAS.404..475J, 2010A&A...520A..43O, 2012A&A...544A..32L, 2012A&A...546A..18M}. In the latter scenario the cores can grow on time scales of $10^6$ years at a radial distance of $5$~AU from the host star \citep{2014arXiv1408.6094L}. Such fast growth depends on the point where a core grows large enough to enter a phase of rapid accretion (so-called Hill accretion), which occurs earlier in regions with lower scale heights.

{\it Planetary migration} describes the gravitational interactions of planets and planetary cores with the gas disc \citep{1997Icar..126..261W}. In a locally isothermal disc, cores are expected to migrate inwards on time scales that are much shorter than the discs' lifetime \citep{2002ApJ...565.1257T}, posing a problem for the formation of the cores of giant planets. However, the migration of cores cores depends on the thermodynamics in the disc, and even outward migration of planetary cores is possible \citep{2006A&A...459L..17P, 2008A&A...487L...9K, 2008ApJ...672.1054B, 2009A&A...506..971K}. The migration depends on the local radial gradient of entropy, which can drive outward migration (see \citet{2013arXiv1312.4293B} for a review).

The structure of the protoplanetary disc that surrounded our own Sun can be approximated by the Minimum Mass Solar Nebular (MMSN) which comes from fitting the solid mass (dust and ice) of the existing planets in our solar system with a power law \citep{1977Ap&SS..51..153W, 1981PThPS..70...35H}. It is then often assumed that other protoplanetary discs have similar power law structures \citep{2013MNRAS.431.3444C}, however applying it to all extrasolar systems is troublesome \citep{2014MNRAS.440L..11R}. Along with the MMSN model, \citet{2010AREPS..38..493C} propose a model that features slightly different gradients in the disc (hereafter named CY2010). Both models, the MMSN and the CY2010 model, approximate the disc structure by uniform power laws in temperature $T$ and gas surface density $\Sigma_G$. The difference in the two models originates in updated condensate mass fractions of solar abundances \citep{2003ApJ...591.1220L} that lead to a higher estimate of the surface density. The different temperature profile is explained by different assumptions on the grazing angle of the disc that determines the absorption of stellar irradiation. In the CY2010 model that follows the calculations of \citep{1997ApJ...490..368C}, $T(r) \propto r^{-3/7}$, in contrast to the MMSN where $T(r) \propto r^{-1/2}$ stemming from the assumption of an optically thin disc.

However, simulations including radiative cooling and viscous and stellar heating have shown that the disc structure features bumps and dips in the temperature profile (\citet{2013A&A...549A.124B} - hereafter Paper I). \citet{2014A&A...564A.135B} (hereafter Paper II) highlighted the direct link in accretion discs between changes in opacity $\kappa$ and the disc profile. The mass flux $\dot{M}$ is defined for a steady state disc with constant $\dot{M}$ at each radius $r$ as
\begin{equation}
\label{eq:Mdotvr}
 \dot{M} = - 2\pi r \Sigma_G v_r \ .
\end{equation}
Here $\Sigma_G$ is the gas surface density and $v_r$ the radial velocity. Following the $\alpha$-viscosity approach of \citet{1973A&A....24..337S} we can write this as
\begin{equation}
 \label{eq:mdot}
 \dot{M} = 3 \pi \nu \Sigma_G = 3 \pi \alpha H^2 \Omega_K \Sigma_G \ .
\end{equation}
Here $H$ is the height of the disc and $\Omega_K$  the Keplerian rotation frequency. At the ice line, the opacity changes because of the melting and the sublimation of ice grains or water vapour. This change in opacity changes the cooling rate of the disc [$D \propto 1 / (\rho \kappa)$], hence changing the temperature in this region. A change in temperature is directly linked to a change in $H$, so that the viscosity changes. However, since the disc has a constant $\dot{M}$ rate at all radii, a change in viscosity has to be compensated by a change in surface density, creating bumps and dips in the disc profile, which do not exist in the MMSN and CY2010 model.

We present detailed 2D ($r,z$) simulations of discs with constant $\dot{M}$, including stellar and viscous heating, as well as radiative cooling. We use the $\alpha$ approach to parametrize viscous heating, but the protoplanetary disc is not temporally evolved with the $\alpha$ approach. Instead $\alpha$ is used to break the degeneracy between column density $\Sigma_G$ and viscous accretion speed $v_r$ (eq.~\ref{eq:Mdotvr}). Each value of $\dot{M}$ is linked to an evolutionary time through observations \citep{1998ApJ...495..385H}. This then allows us to take the correct stellar luminosity from stellar evolution \citep{1998A&A...337..403B}. In contrast to previous work \citep{2014A&A...570A..75B}, we do not include transitions in the $\alpha$ parameter to mimic a dead zone, because we are primarily interested in investigating how the stellar luminosity and the disc's metallicity affect the evolution of the disc structure in time. For the time evolution of the disc over several Myr (and therefore several orders of magnitudes in $\dot{M}$), we present a semi-analytical formula that expresses all disc quantities as a function of $\dot{M}$ and metallicity $Z$.

This paper is structured as follows. In section~\ref{sec:methods} we present the numerical methods used and the link between $\dot{M}$ and the time evolution. We point out important differences in the disc structure between our model and the MMSN and CY2010 model in section~\ref{sec:discstructure}. We show the influence of the temporal evolution of the star on the structure of discs with different $\dot{M}$ in section~\ref{sec:timeevolve}. We then discuss the influence of metallicity on the disc structure and evolution in section~\ref{sec:metallicity}. We then discuss the implications of the evolution of protoplanetary discs on planet formation in section~\ref{sec:formplanets}. We finally summarize in section~\ref{sec:summary}. In appendix~\ref{ap:model} we present the fits for the full time evolution model of protoplanetary discs.

\section{Methods}
\label{sec:methods}

\subsection{Numerical set-up}

We treat the protoplanetary disc as a three-dimensional (3D), non-self-gravitating gas, whose motion is described by the Navier-Stokes equations. We assume an axisymmetric disc structure because we do not include perturbers (e.g. planets) in our simulations. Therefore we use only one grid cell in the azimuthal direction, making the computational problem de facto 2D in the radial and vertical direction. We utilize spherical coordinates ($r$-$\theta$) with $386 \times 66$ grid cells. The viscosity is treated in the $\alpha$-approach \citep{1973A&A....24..337S}, where our nominal value is $\alpha=0.0054$. Here the viscosity is used as a heating parameter and not to evolve the disc viscously, because the viscous evolution of the disc is very slow. Instead we use an initial radial gas density profile from a 1D analytic model for each accretion rate $\dot{M}$. The vertical profile is computed from an analytic model of passive discs irradiated by the central star. The simulations are then started until they reach an equilibrium state between heating and cooling, which happens much faster than the viscous evolution of the disc. This final equilibrium state is different from the initial one, because the disc is not passive (viscous heating is included) and opacities depend on the temperature.
This changes $H/r$ in the disc, which in turn changes the local viscosity because of the $\alpha$-prescription. Therefore we continue the simulations until a new radial density profile in the disc is achieved. This happens on a viscous time scale, but because the changes are only local variations relative to the initial profile, the new equilibrium state is achieved relatively fast, much faster than the global evolution of the disc and decay of the stellar accretion rate (see section~\ref{subsec:time}).

The viscosity in protoplanetary discs can be driven by the magnetorotational instability (MRI), where ionized atoms and molecules cause turbulence through interactions with the magnetic field \citep{1998RvMP...70....1B}. As ionization is more efficient in the upper layers of the disc (thanks to cosmic and X-rays), the midplane regions of the disc are not MRI active and therefore feature a much smaller viscosity ($\alpha$ parameter). In discs with a constant $\dot{M}$, a change in viscosity has to be compensated for by an equal change in surface density (eq.~\ref{eq:mdot}); however, in 3D simulations, much of the accretion flow can be transported through the active layer, so that the change in surface density is smaller than for 2D discs \citep{2014A&A...570A..75B}. Additionally, hydrodynamical instabilities, such as the baroclinic instability \citep{2003ApJ...582..869K} or the vertical shear instability \citep{2013MNRAS.435.2610N, 2014arXiv1409.8429S}, can act as a source of turbulence in the weakly ionised regions of the disc. A realistic picture of the source of turbulence inside accretion discs is still being debated (see e.g. \citet{Turner2014}). We therefore feel that it is legitimate to neglect the effects of a dead zone and assume a constant $\alpha$ throughout the disc.

The dissipative effects can then be described via the standard viscous stress-tensor approach \citep[e.g.][]{1984frh..book.....M}. We also include irradiation from the central star, as described in Papers I and II. For that purpose we use the multi-dimensional hydrodynamical code FARGOCA, as originally presented in \citet{2014MNRAS.440..683L} and in Paper II. The radial extent of our simulations spans from $1$ AU to $50$ AU, which includes the range of the MMSN, which is defined from $0.4$ to $36$ AU. We apply the radial boundary conditions described in the appendix of Paper II.

The radiative energy associated with viscous heating and stellar irradiation is transported through the disc and emitted from its surfaces. To describe this process we utilize the flux-limited diffusion approximation \citep[FLD,][]{1981ApJ...248..321L}, an approximation that allows the transition from the optically thick mid-plane to the thin regions near the disc's surface.

The hydrodynamical equations solved in the code have already been described in detail \citep{2009A&A...506..971K}, and the two-temperature approach for the stellar irradiation was described in detail in Paper I, so we refrain from quoting it here again. The flux $F_\star$ from the central star is given by
\begin{equation}
\label{eq:luminosity}
 F_\star = \frac{R_\star^2 \sigma T_\star^4}{r^2} \ ,
\end{equation}
where $R_\star$ and $T_\star$ give the stellar radius and temperature and $\sigma$ is the Stefan–Boltzmann constant. Stellar heating is responsible for keeping the disc flared in the outer parts (Paper I). The size and temperature of the star changes in time (see section~\ref{subsec:time}) and the corresponding values are displayed in table~\ref{tab:Starsize}, where $t_\star$ gives the age of the star. The stellar mass is fixed to $1 M_\odot$. We describe the effects of time on the size and temperature of the star in section~\ref{subsec:time} and take these effects into account in section~\ref{sec:timeevolve}.

 {%
 \begin{table}
  \centering
  \begin{tabular}{ccccc}
  \hline \hline
  $\dot{M}$ in $M_\odot/\text{yr}$ & $T_\star$ in K & $R_\star$ in $R_\odot$ & $L$ in $L_\odot$ & $t_\star$ in Myr  \\ \hline
  {$1 \times 10^{-7}$} & 4470 & 2.02 & 1.47 & 0.20 \\
  {$7 \times 10^{-8}$} & 4470 & 1.99 & 1.42 & 0.25 \\
  {$3.5 \times 10^{-8}$} & 4450 & 1.92 & 1.31 & 0.41  \\
  {$1.75 \times 10^{-8}$} & 4430 & 1.83 & 1.16 & 0.67 \\
  {$8.75 \times 10^{-9}$} & 4405 & 1.70 & 0.98 & 1.10 \\
  {$4.375 \times 10^{-9}$} & 4360 & 1.55 & 0.80 & 1.80 \\
  \hline
  \end{tabular}
  \caption{Stellar parameters for different $\dot{M}$ as time evolves.
  \label{tab:Starsize}
  }
 \end{table}
 }%

The initial surface density profile follows $\Sigma_G \propto r^{-15/14}$, which follows from eq.~\ref{eq:mdot} for a flared disc with $H/r \propto r^{2/7}$. We model different $\dot{M}$ values by changing the underlying value of the surface density, while we keep $\alpha$ constant. This does not imply that the same viscosity ($\nu = \alpha H^2 \Omega$) is present at all the different $\dot{M}$ stages, because $H$ changes as the disc evolves (Paper II). In our simulations we set the adiabatic index $\gamma = 1.4$ and used a mean molecular weight of $\mu =2.3$.

We use the opacity profile of \citet{1994ApJ...427..987B}, which is derived for micrometer-sized grains. In fact dust growth can be quite fast \citep{2010A&A...513A..57Z}, depleting the micro meter-sized dust grains to some level. However, larger grains (starting from mm size) only make a very small contribution to the opacity at high temperatures. At lower temperatures ($T < 15$ K), the larger mm grains will dominate the opacity, but those temperatures are not relevant within $50$AU. Additionally, if these grains start to grow and form larger pebbles, they will make a zero contribution to the opacity. 

We define $\Sigma_Z$ as the surface density of heavy elements that are $\mu m$ size in condensed form and $\Sigma_G$ as the gas surface density. Thus, the metallicity $Z$ is the ratio $Z=\Sigma_Z / \Sigma_G$, assumed to be independent of $r$ in the disc. We assume that the grains are perfectly coupled to the gas, meaning that the dust-to-gas ratio is the same at every location in the disc. In our simulations we assume metallicities from $0.1\%$ to $3.0\%$. This means that if grain growth occurs and the total amount of heavy elements (independent of size) stays the same, then the metallicity in our sense is reduced (as in Paper II).

\subsection{Time evolution of discs}
\label{subsec:time}

Protoplanetary accretion discs evolve with time and reduce their $\dot{M}$. Observations can give a link between mass accretion $\dot{M}$ and time $t$. \citet{1998ApJ...495..385H} find a correlation between the mass accretion rate and the star age
\begin{equation}
\label{eq:harttime}
 \log \left( \frac{\dot{M}}{M_\odot /\text{yr}} \right) = -8.00 \pm 0.10 - (1.40 \pm 0.29) \log \left( \frac{t}{10^6 \text{yr}} \right) \ .
\end{equation}
This correlation includes stars in the Taurus star cluster that span an $\dot{M}$ range from $\dot{M}=5 \times 10^{-6} M_\odot$/yr to $\dot{M}=5 \times 10^{-10} M_\odot$/yr over a range of $10$ Myr. If the disc is accreting viscously, the evolution of the disc is directly proportional to the viscosity and hence to $\alpha$. However, eq.~\ref{eq:harttime} was derived without any parametrization in $\alpha$. We therefore consider that $\alpha$ in our simulations is not the time evolution parameter of the surface density, but simply a parameter for viscous heating ($Q^+ \propto \alpha$) and for determining $v_r$. The time evolution of the disc is parametrized by eq.~\ref{eq:harttime}, where we use the values without the error bars for our time evolution.

This approach also implies that the evolution time of the disc has to be longer than the time to relax to a radially constant $\dot{M}$ state. This is more critical in the early evolution of the disc (high $\dot{M}$), because the disc evolves more rapidly at that point. But the time the simulations need to settle in a steady state (constant $\dot{M}$ at all $r$) is about a factor of a few ($\approx 5$) shorter than the disc evolution time in eq.~\ref{eq:harttime} for high $\dot{M}$ values and is much shorter for the small $\dot{M}$ ranges, validating our approach.  

A disc with $\dot{M}=1 \times 10^{-9} M_\odot$/yr can be cleared by photo evaporation quite easily, so that the remaining lifetime is very short, only a few thousand years \citep{2013arXiv1311.1819A}. Therefore we do not simulate discs with very low $\dot{M}$.
Recent observations have reported that even objects as old as $10$ Myr can have accretion rates up to $\dot{M}=1 \times 10^{-8} M_\odot$/yr \citep{arXiv:1406.0722}, which is in contrast to eq.~\ref{eq:harttime}. Such high accretion rates at that age cannot be explained by viscous evolution models, unless the disc was very massive in the beginning of the evolution. Even if more of these objects are observed, this will not change the validity of the approach we are taking here. It would just change the time evolution of the disc presented in eq.~\ref{eq:harttime}, but the disc structure as a function of $\dot{M}$ (presented in section~\ref{sec:timeevolve}) would stay the same.

\begin{figure}
 \centering
 \includegraphics[scale=0.71]{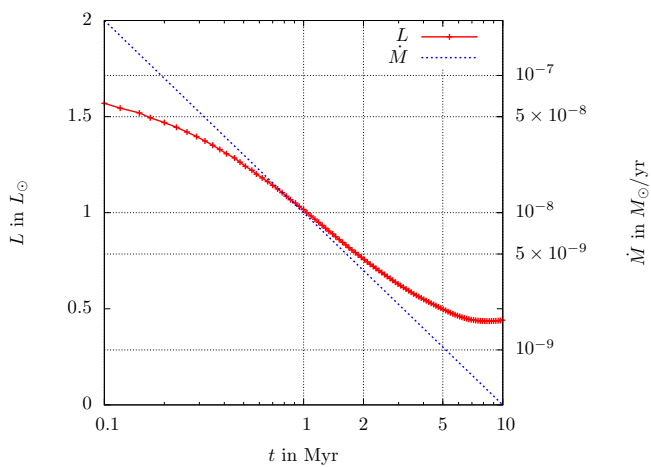}
 \caption{Time evolution of the stellar luminosity after \citet{1998A&A...337..403B} and the evolution of $\dot{M}$ after \citet{1998ApJ...495..385H}. The luminosity of the star reduces by a factor of $3$ in $10$ Myr and the accretion rate reduces by over $2$ orders of magnitude during the same time period. However, an accretion rate of $\dot{M}=1 \times 10^{-9} M_\odot$/yr is already reached after $5$ Myr when the disc can be cleared by photo evaporation.
   \label{fig:LMdot} 
   }
\end{figure}

During the millions of years of the disc's evolution, the star evolves as well \citep{1998A&A...337..403B}. As the star changes temperature and size, its luminosity changes (eq.~\ref{eq:luminosity}), influencing the amount of stellar heating received by the disc. The stellar evolution sequence used was calculated by \citet{1998A&A...337..403B} (see table~\ref{tab:Starsize}), where we display the stellar age, temperature, density, luminosity, and the corresponding accretion rate $\dot{M}$ from \citet{1998ApJ...495..385H}. In Fig.~\ref{fig:LMdot} we plot the stellar luminosity and the $\dot{M}$ rate of the disc as a function of time. As the stellar luminosity drops by a factor of $3$, the $\dot{M}$ rate decreases by two orders of magnitude. Summarizing our methods, we have a disc model that features the full 2D structure with realistic heating and cooling that is linked to the temporal evolution of the star and disc. 

The total mass flowing through the disc can be evaluated by integrating the accretion rate specified in eq.~\ref{eq:harttime} over time. This gives a minimum estimate of the total mass of the disc, as leftover material is blown away by photo evaporation as soon as $\dot{M}<1 \times 10^{-9} M_\odot$/yr. During the disc's lifetime of $5$ Myr, a total of $0.05 M_\odot$ of gas flows through the disc, marking the total mass of the disc.

We therefore present fits to our simulations in Appendix~\ref{ap:model} that can then be easily used by other studies that need a simple, but accurate time evolution model of the accretion disc structure.

\section{Disc structure}
\label{sec:discstructure}

In this section we compare the structure of a simulated $\dot{M}$ disc with the MMSN and the CY2010 nebula to point out crucial differences in the disc structure and their effect on the formation of planetesimals and planetary embryos. The simulations in this section feature a metallicity of $0.5\%$ in $\mu m$-sized dust grains. This value allows the disc to contain more heavy elements that could represent pebbles, planetesimals, or planetary cores (that do not contribute to the opacity) without increasing the total amount of heavy elements (grains and larger particles) to very high values.

\subsection{$\dot{M}$ disc}
\label{subsec:dotMdisc}

\begin{figure}
 \centering
 \includegraphics[scale=0.71]{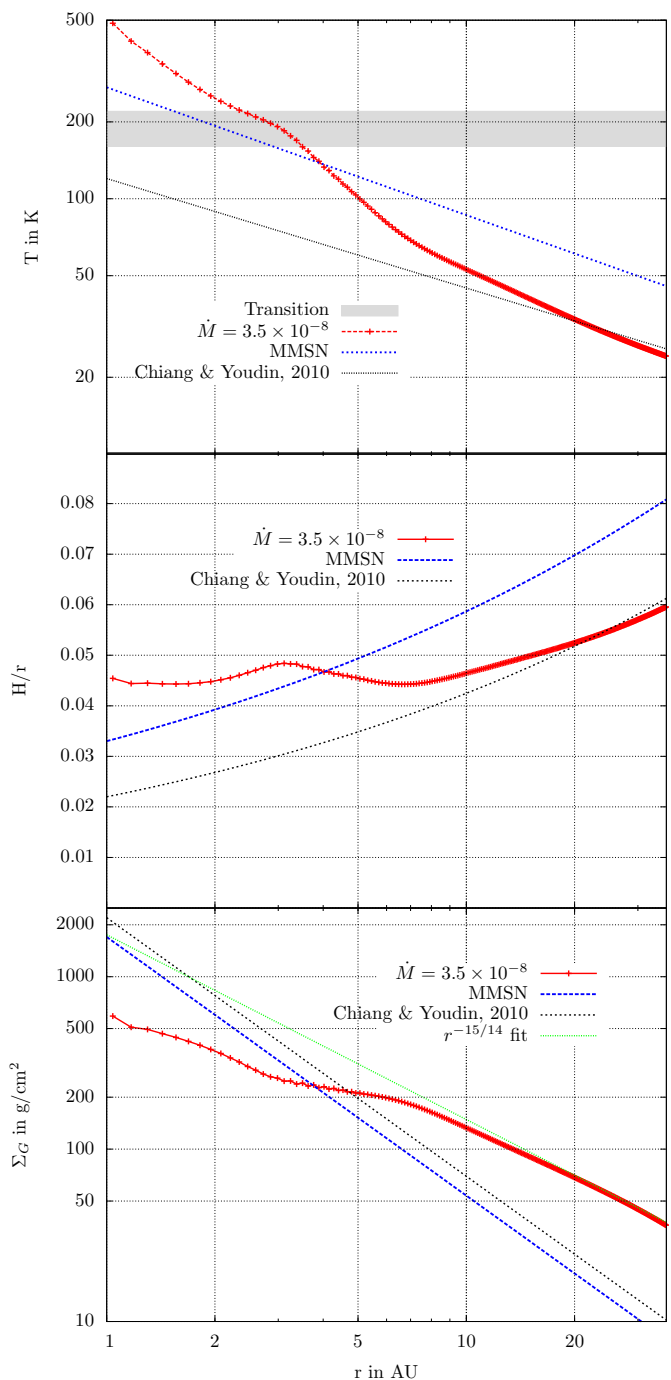}
 \caption{Mid-plane temperature $T$ (top), aspect ratio $H/r$ (middle) and integrated surface density $\Sigma_G$ (bottom) for a disc with $\dot{M}= 3.5 \times 10^{-8} M_\odot$/yr and for the MMSN and the CY2010 model. The grey area marks in the temperature plot the region of the opacity transition at the ice line. The green line in the surface density plot indicates a fit for the surface density in the outer part of the disc. The main difference between the $\dot{M}$ disc and the MMSN and CY2010 model is the higher temperature in the inner parts of the disc, which is caused by viscous heating. This increases $H/r$ in the inner parts and then results in a reduced $\Sigma_G$ in the inner parts, owing to a change in viscosity that is compensated by a change in $\Sigma_G$. 
   \label{fig:HrTSigall} 
   }
\end{figure}

In Fig.~\ref{fig:HrTSigall} the temperature (top), the aspect ratio $H/r$ (middle), and the surface density $\Sigma_G$ (bottom) are displayed. The MMSN and the CY2010 model follow power laws in all disc quantities. These are quoted in table~\ref{tab:power}. The simulated $\dot{M}=3.5 \times 10^{-8} M_\odot$/yr disc model features bumps and dips in all disc quantities. More specifically, the simulation features a bump in $T$ at the ice line (at $T_{\text{ice}} \approx 170$ K, illustrated in Fig.~\ref{fig:HrTSigall}). As ice grains sublimate at higher temperature, the opacity reduces for larger $T$ and therefore the cooling rate of the disc [$D \propto 1/(\kappa \rho)$] increases. This reduces the gradient of the temperature for high $T$ compared to $T<T_{\text{ice}}$, creating an inflection in $T$. This also changes the scale height $H/r$ of the disc, which in turn influences the viscosity. Since the disc features a radially constant $\dot{M}$, the disc adapts to this change in viscosity by changing $\Sigma_G$, creating the flattening of $\Sigma_G$ at the ice line (Paper II).

{%
\begin{table}
 \centering
 \begin{tabular}{ccc}
 \hline \hline
  & \textbf{MMSN} & \textbf{CY2010} \\\hline
 {\textbf{$H/r$}} & 1/4 & 2/7 \\
 {\textbf{$T$}} & -1/2 & -3/7 \\
 {\textbf{$\Sigma_G$}} & -3/2 & -3/2  \\
 {\textbf{$\Delta v / c_s$}} & 1/4 & 2/7  \\
 \hline
 \end{tabular}
 \caption{Parameters of the power laws used in the MMSN and the CY2010 models.
 \label{tab:power}
 }
\end{table}
}%

In the inner parts of the disc, our models feature a much higher temperature than in the MMSN and the CY2010 models due to the inclusion of viscous heating. In the outer parts of the disc, where viscous heating becomes negligible and the temperature is solely determined by stellar irradiation, the temperatures of our simulations are comparable to the CY2010 model. 

The aspect ratio $H/r$ is related quite simply to the temperature (through the relation of hydrostatic equilibrium);
\begin{equation}
 T = \left( \frac{H}{r} \right)^2 \frac{G M_\star}{r} \frac{\mu}{\cal R} \ ,
\end{equation}
where $\cal R$ is the gas constant, $G$ the gravitational constant, and $M_\star$ the mass of the star. Therefore, the $H/r$ plot (middle panel in Fig.~\ref{fig:HrTSigall}) features the same properties as the temperature plot: i) a higher $H/r$ in the inner disc for our simulations and ii) about the same $H/r$ in the outer disc compared to the CY2010 model. Strikingly, the aspect ratio of the MMSN is off by about $50\%$ at $20$ AU compared to our simulations. As the MMSN model does not feature viscous heating in the inner parts, $H/r$ is smaller there, and because the radial change of $H/r$ is roughly the same in both models in the outer parts of the disc, the MMSN has a much smaller $H/r$ in the outer parts of the disc. The $H/r$ diagram now features bumps and dips correlated to the wiggles in the temperature diagram. The dip in $H/r$ starting beyond $3$ AU represents a shadowed regions inside the disc. The stellar irradiation does not penetrate well into this region because it is absorbed by the bump in $H/r$ at $\approx 3$ AU. We emphasise that the drop in $H/r$ is caused by the change of the cooling rate in the disc as the opacity changes and not by the heat transition from regions dominated by viscous heating to regions dominated by stellar heating (Paper II). If a constant opacity had been used, shadowed regions would not have appeared in the disc (Paper I). A drop in $H/r$ normally implies outward migration for low mass planets (Papers I and II), a phenomenon that is not seen in the MMSN and CY2010 models.

The surface density profile (bottom panel in Fig.~\ref{fig:HrTSigall}) of our simulations shows an inflection at the same location as where $H/r$ shows a bump. Generally our simulations feature a lower surface density in the inner parts of the disc than in the MMSN and the CY2010 nebula. In the outer parts, on the other hand, the surface density of our simulations is much higher. This difference in surface density between our simulations and the other two models is caused by our underlying $\dot{M}$ approximation (see eq.~\ref{eq:mdot}), which gives a shallower surface density slope for the outer disc. \citet{2010ApJ...723.1241A} find that the gradients of the surface density profile of accretion discs in the Ophiuchus star forming region are between $0.4$ and $1.1$. Our simulations feature a surface density gradient of $\approx 1$ in the outer parts of the disc, but it is much shallower in the inner parts, matching the observations in contrast to the MMSN and CY2010 model.

\subsection{Influence on planet formation}
\label{subsec:planetform}

The streaming instability can lead to formation of planetesimals as a first step in forming planetary embryos and planets \citep{2007Natur.448.1022J,2007ApJ...662..627J,2010ApJ...722.1437B,2010ApJ...722L.220B}. The important quantity for triggering particle concentration by the streaming instability is $\Delta$, which is the difference between the azimuthal mean gas flow and the Keplerian orbit divided by the sound speed. This difference is caused by the reduction of the effective gravitational force by the radially outwards pointing force of the radial pressure gradient. The parameter $\Delta$ is given by
\begin{equation}
\label{eq:stream}
 \Delta = \frac{\Delta v}{c_s} = \eta \frac{v_K}{c_s} = - \frac{1}{2} \frac{H}{r} \frac{\partial \ln (P)}{\partial \ln (r)} \ ,
\end{equation}
where $v_K=\sqrt{GM/r}$ is the Keplerian velocity and $c_s / v_K = H/r$. Here $\eta$ represents a measure of the gas pressure support \citep{1986Icar...67..375N}. The sub-Keplerian rotation of the gas makes small solid particles drift towards the star. 

\begin{figure}
 \centering
 \includegraphics[scale=0.71]{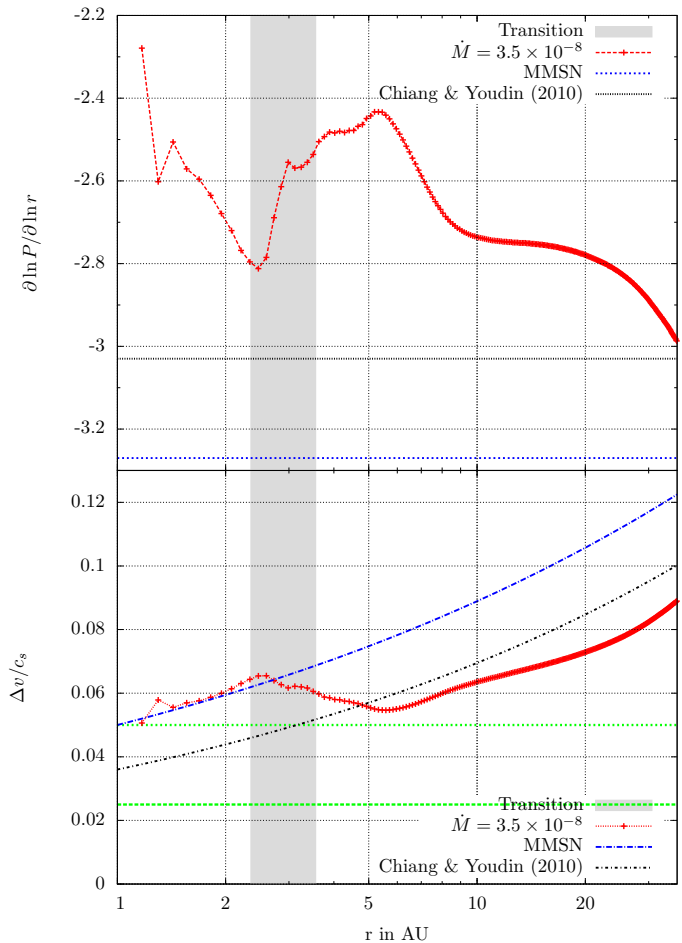}
 \caption{Radial pressure gradient (top) and pressure support parameter $\Delta = \Delta v/c_s$ (bottom) for the disc with $\dot{M}= 3.5 \times 10^{-8} M_\odot$/yr and for the MMSN and CY2010 models. The grey area indicates the radial range of the opacity transition at the ice line, as already indicated in Fig.~\ref{fig:HrTSigall}. In contrast to the MMSN and CY2010 models that have a constant pressure gradient and therefore a steadily increasing $\Delta$ parameter, the $\dot{M}$ model features a dip in the profile at $\approx 5$ AU, which makes the formation of planetesimals in this region more likely. The reduced $\Delta$ parameter in the outer parts of the disc also makes it more likely to form planetesimals there compared to the MMSN and CY2010 model. The horizontal green lines at $\Delta=0.025$ and $\Delta=0.05$ mark the amount of heavy elements in the disc needed for the streaming instability to operate ($\approx 1.5\%$ and $\approx 2\%$, respectively), see \citep{2010ApJ...722L.220B}.
   \label{fig:Petaall}
   }
\end{figure}

In Fig.~\ref{fig:Petaall} the radial pressure gradient $\partial \ln(P) / \partial \ln(r)$ (top) and $\Delta$ (bottom) are displayed. The radial pressure gradient is constant for the MMSN and the CY2010 model, since these models are built upon strict power laws. The simulations, on the other hand, feature bumps and dips that result from the opacity transition. In the inner parts of the disc, the pressure gradient is much more shallow in the $\dot{M}$ disc. We recall here that a shallower negative pressure gradient is preferred in order to form planetesimals. 

The $\Delta$ parameter (bottom in Fig.~\ref{fig:Petaall}) is nearly constant for our simulation in the inner parts of the disc, while it steadily decreases towards the star in the MMSN and CY2010 models. In the outer parts, $\Delta$ is up to $50\%$ smaller in the $\dot{M}$ model than in the MMSN, indicating significantly different conditions for forming planetesimals.

A reduced $\Delta$ parameter significantly helps the formation of large clumps via the streaming instability \citep{2010ApJ...722L.220B}. Additionally, the formation of clumps is also dependent on the amount of heavy elements and particle sizes in the disc, where a larger amount of heavy elements strongly increases the clumping. The number of heavy elements is not restricted to the metallicity defined with $\mu m$ dust grains, but also includes larger grains and pebbles that do not contribute to the opacity profile of the disc. For $\Delta=0.025$  (lower horizontal line in Fig.~\ref{fig:Petaall}), a fraction of $\approx 1.5\%$ in heavy elements is needed to form large clumps, while already for $\Delta=0.05$ (top horizontal green line in Fig.~\ref{fig:Petaall}), a fraction of $\approx 2\%$ in heavy elements is needed. For $\Delta = 0.1$ a very high, probably not achievable, number of heavy elements is needed for the streaming instability to work. A reduction in $\Delta$ in the outer parts of the disc, as proposed by our model, therefore makes the formation of planetesimals much easier in the Kuiper belt.

Planetesimals can grow further by mutual collisions \citep{2010AJ....139.1297L} or by the accretion of pebbles. In the latter case, core growth enters the fast Hill regime, when it reaches the `transition mass', 
\begin{equation}
\label{eq:Mtrans}
M_{\rm t} \approx \sqrt{\frac{1}{3}} \Delta^3 \left( \frac{H}{r} \right)^3 M_\star
\end{equation}
where $M_\star$ is the stellar mass \citep{2012A&A...544A..32L}. This corresponds to $0.03 M_{\rm Earth}$ at $5.2$ AU in our simulated $\dot{M}$ disc. A reduced disc scale height and $\Delta$ parameter thus help smaller embryos to reach this growth regime where cores are formed on time scales of $10^5$\,yr, even at wide orbits (50\,AU), provided the pebble surface density is similar to MMSN estimates.  Furthermore, a lower $\Delta$ makes accretion more efficient by increasing the proportion of pebbles that are accreted by a core versus those particles that drift past \citep{2012ApJ...747..115O}.

Once planetary embryos have formed, these embryos are subject to gas-driven migration (for a review see \citet{2013arXiv1312.4293B}). The migration rate of planets can be determined by 3D simulations of protoplanetary discs \citep{2009A&A...506..971K, 2011A&A...536A..77B}. However, these simulations are quite computationally expensive. Instead, one can calculate the torque acting on an embedded planet from the disc structure \citep{2010MNRAS.401.1950P, 2011MNRAS.410..293P}. The torque formula of \citet{2011MNRAS.410..293P} takes torque saturation due to thermal diffusion into account and matches the 3D simulations of planets above $15M_{\text{Earth}}$  well\citep{2011A&A...536A..77B}. However, for low mass planets ($M_P \leq 5M_{\text{Earth}}$), there is still a discrepancy between the torque formula and 3D simulations that actually show a more negative torque \citep{2014MNRAS.440..683L}. Nevertheless, the predictions from the torque formula can already give first clues about the planetary migration history in evolving accretion discs.

In \citet{2011MNRAS.410..293P} the total torque acting on an embedded planet is a composition of its Lindblad torque and its corotation torque
\begin{equation}
 \Gamma_{\text{tot}} = \Gamma_{\text L} + \Gamma_{\text C} \ .
\end{equation}
The Lindblad and corotation torques depend on the local radial gradients of surface density $\Sigma_G \propto r^{-s}$, temperature $T \propto r^{-\beta}$, and entropy $S \propto r^{-\xi}$, with $\xi = \beta - (\gamma - 1.0) s$. Very roughly said, for not too negative $\Sigma_G$ gradients, a radially strong negative gradient in entropy, caused by a large negative gradient in temperature (large $\beta$), will lead to outward migration, while a shallow gradient in entropy will not lead to outward migration.

\begin{figure}
 \centering
 \includegraphics[width=1.0\linwx]{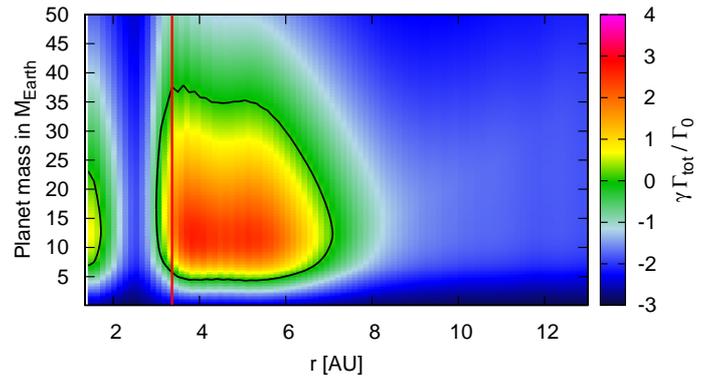}
 \caption{Torque acting on discs with $\dot{M}=3.5 \times 10^{-8} M_\odot$/yr. The black lines encircle the regions of outward migration. The vertical red lines indicate the ice line at $170$K. The region of outward migration is exactly correlated to the region in the disc, where the temperature gradient is the steepest and can trap planets between $5$ and $30$ Earth masses.
   \label{fig:Migsim}
   }
\end{figure}

In Fig.~\ref{fig:Migsim} the migration maps for the $\dot{M}=3.5 \times 10^{-8} M_\odot$/yr disc is displayed. The torque parameter $\Gamma_0$ is defined as
\begin{equation}
 \Gamma_0 = \left(\frac{q}{h}\right)^2 \Sigma r^4 \Omega^2 \ ,
\end{equation}
where $q$ is the planet-to-star mass ratio, $h=H/r$, $r$ the semi-major axis and $\Omega_P$ the Keplerian rotation. The actual speed of inward migration changes with planetary mass not only because the torque is proportional to the planetary mass squared, but also because the mass changes the saturation effects of the corotation torque \citep{2011MNRAS.410..293P}.

The regions of outward migration correspond to the regions in the disc where $H/r$ decreases with $r$. The MMSN and the CY2010 model feature a flared disc for all $r$ (Fig.~\ref{fig:HrTSigall}). This imposes a very shallow temperature gradient leading to a shallow entropy gradient, which is not enough to produce a corotation torque that can overcompensate the Lindblad torque. Planets, regardless of mass, therefore migrate inwards towards the star in both the MMSN and CY2010 models. We do not display these migration maps here. This lack of zones of outward migration makes the formation of giant planet cores much harder.

\section{Influence of the time evolution of disc and star}
\label{sec:timeevolve}

The protoplanetary disc gets accreted onto the star over millions of years, but the star also evolves on these time scales \citep{1998A&A...337..403B}, contracting and changing its size and luminosity. The luminosity in turn determines the stellar heating that is absorbed by the disc, and thus influences the disc structure. By using eq.~\ref{eq:harttime}, we can link different $\dot{M}$ stages to different times. These times then give us the stellar evolution time and thus the stellar radius and temperature (table~\ref{tab:Starsize}). The stellar temperature stays roughly constant with time, while the stellar radius becomes significantly smaller as time evolves. This means that for discs with lower $\dot{M}$, the stellar heating will decrease as well. This will influence the outer regions of the disc, which are dominated by stellar heating. An increase of a factor of $2$ in stellar luminosity results in a change of $H/r$ of up to $\approx 20\%$ and of up to $\approx 45\%$ in temperature in the parts dominated by stellar irradiation. In this section all shown simulations feature a metallicity of $0.5\%$, as in section~\ref{sec:discstructure}.

\begin{figure}
 \centering
 \includegraphics[scale=0.71]{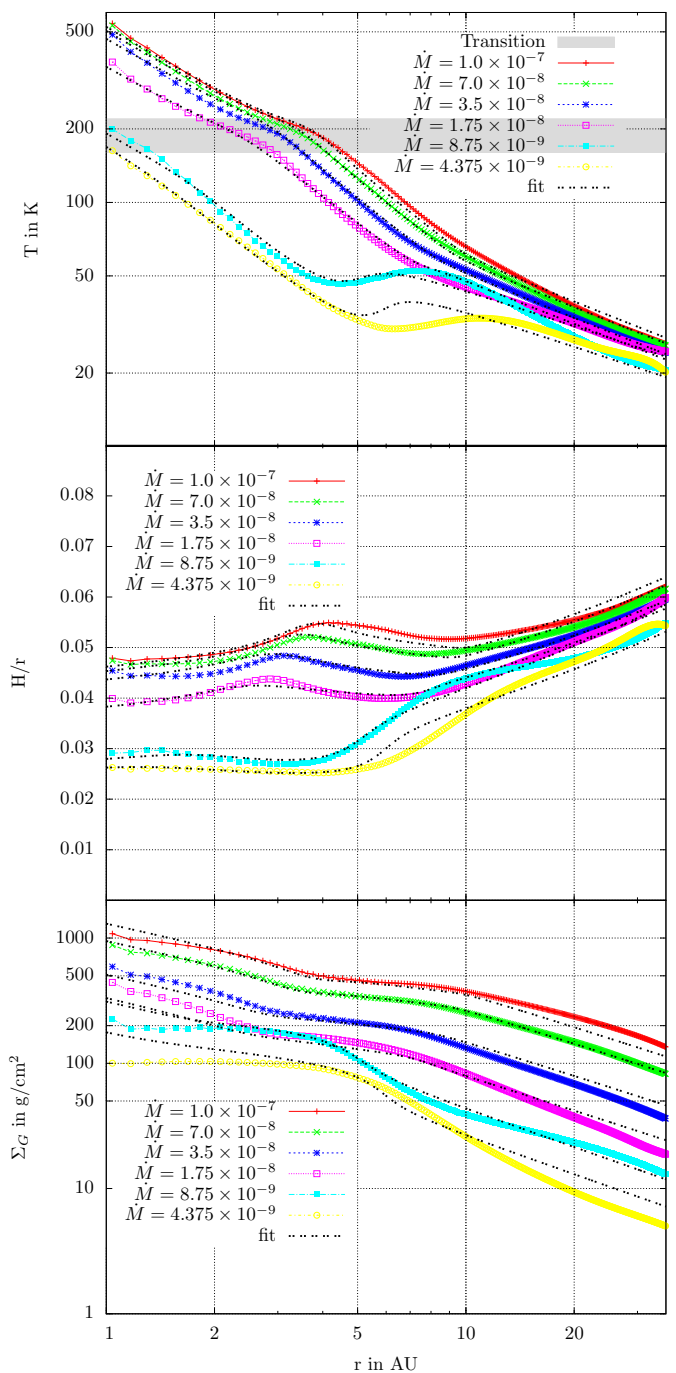}
 \caption{Mid-plane temperature (top), $H/r$ (middle), and surface density $\Sigma_G$ (bottom) for discs with different values of $\dot{M}$ around an evolving star. The grey area in the temperature plot marks the transition range in temperature for the opacity law at the ice line. The black lines mark the fits discussed in Appendix~\ref{ap:model}. For lower $\dot{M}$ rates, the inner parts of the disc are colder, as viscous heating is reduced. For very low $\dot{M}$, the temperature is below the ice condensation temperature throughout the disc's mid plane, and therefore the bump in the inner part of the protoplanetary disc vanishes.
   \label{fig:HrTSigallfit}
   }
\end{figure}

In Fig.~\ref{fig:HrTSigallfit} the mid-plane temperature (top), $H/r$ (middle), and the surface density $\Sigma_G$ (bottom) are displayed for different values of $\dot{M}$ ranging from $\dot{M}=1 \times 10^{-7} M_\odot$/yr to $\dot{M}=4.375 \times 10^{-9} M_\odot$/yr. The temperature in the inner parts of the disc drops as $\dot{M}$ decreases, because the viscous heating decreases as $\Sigma_G$ decreases. In the outer parts of the disc, the temperature also drops, because the stellar irradiation decreases in time. However, this drop in temperature is not as large as in the inner parts of the disc. In the late stages of the disc evolution, the disc becomes so cold that ice grains will exist throughout the disc so that no opacity transition is visible any more. For low $\dot{M}$ there is a temperature inversion at approximately $5$ AU. This inversion is caused by the different vertical heights of the absorptions of stellar photons compared to the vertical height of the cooling through diffusion.

As $\dot{M}$ evolves, the shadowed regions of the disc (seen by a reduction in $H/r$ as a function of $r$) shrink. For very low $\dot{M}$, no bumps in $H/r$ exist any more, because the temperature of the disc is so low that the opacity transition is no longer inside the computed domain, but further inside at $r<1$ AU. This will have important consequences for planet migration, since outward migration is only possible when radially strong negative temperature gradients exist.

The wiggles in the temperature profile directly translate to dips in the surface density profile, because a change in the viscosity of the disc must be directly compensated for by a change in the disc's surface density (eq.~\ref{eq:mdot}) to maintain a constant $\dot{M}$. In the very late stages of the disc evolution, when the accretion rate and the surface density are very low ($\dot{M} \leq 1 \times 10^{-9} M_\odot$/yr), the disc will experience rapid photo-evaporation \citep{2013arXiv1311.1819A}.

\begin{figure}
 \centering
 \includegraphics[scale=0.71]{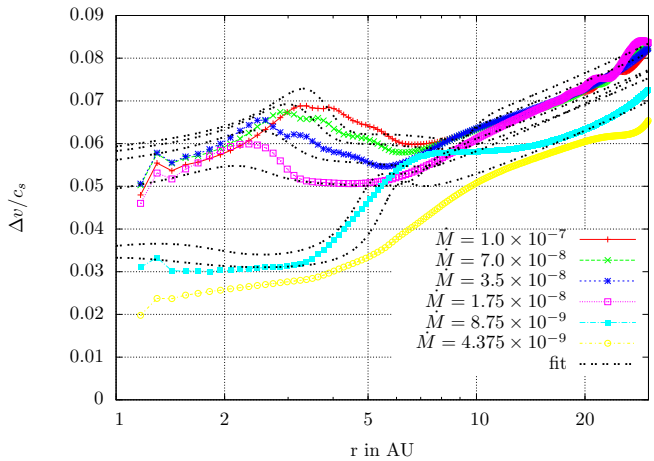}
 \caption{ $\Delta$ parameter for the disc simulations with evolving $\dot{M}$. The black lines mark the fits discussed in Appendix~\ref{ap:model}. The $\Delta$ parameter describes the triggering of particle concentrations in the streaming instability, which can lead to planetesimal formation. In the regions of lower $\Delta$ at $\approx 5$ AU, the formation of planetesimals is more likely, compared to the inner regions of the disc ($\approx 2$ AU) where $\Delta$ is higher.
   \label{fig:etaall}
   }
\end{figure}

In Fig.~\ref{fig:etaall} the $\Delta$ parameter (eq.~\ref{eq:stream}) for the discs with different $\dot{M}$ is displayed. In the inner parts of the disc, $\Delta$ only slightly reduces as $\dot{M}$ shrinks, as long as $\dot{M}$ is still high enough that the inner parts of the disc are dominated by viscous heating. As soon as the disc starts to become dominated by stellar heating in the inner parts, $\Delta$ drops by a significant factor. This is because $\Delta$ is proportional to $H/r$, which shows exactly that behaviour as well. This reduction of $\Delta$ for low $\dot{M}$ helps the formation of planetesimals significantly, since lower metallicity is needed to achieve clumping \citep{2010ApJ...722L.220B}, indicating that in the very late stages of the disc evolution, planetesimal formation becomes easier. In the outer parts of the disc, $\Delta$ is very high throughout the different $\dot{M}$ stages and changes only slightly, following the reduction in $H/r$ as the stellar luminosity decreases. This indicates that for the formation of planetesimals, very high metallicity is needed in the outer disc.

The growth of planetary embryos via pebble accretion is significantly accelerated, when the embryo reaches the Hill accretion regime, defined in eq.~\ref{eq:Mtrans}. Through all $\dot{M}$ stages, the minimum of $H/r$ and $\Delta$ coincides, reducing the transition mass towards the Hill accretion regime and making planet formation in these locations easier. Additionally, as $H/r$ and $\Delta$ drop in the late stages of evolution, pebble accretion is much more efficient because the transition mass is reduced, making core growth more efficient in the late stages of the disc, provided that enough pebbles are still available in the disc.

\begin{figure}
 \centering
 \includegraphics[width=1.0\linwx]{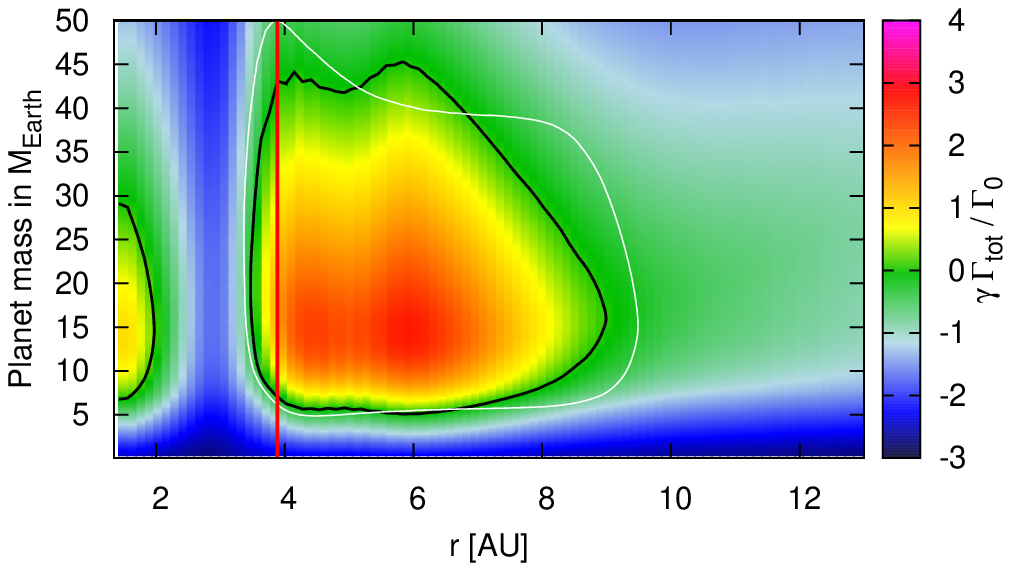}
 \includegraphics[width=1.0\linwx]{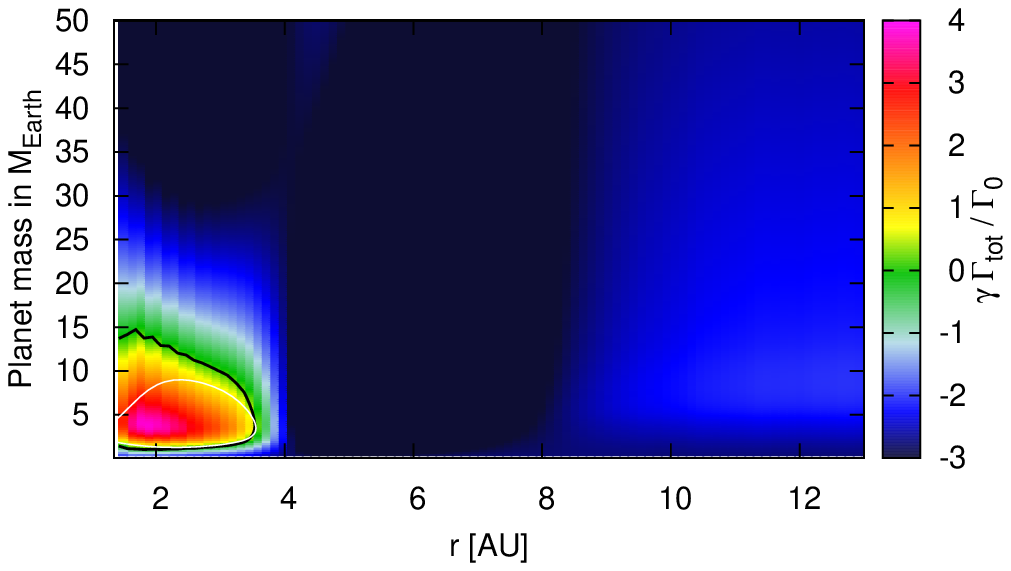}
 \caption{Gravitational torque acting on planets in discs with $\dot{M}=7 \times 10^{-8} M_\odot$/yr (top) and $\dot{M}= 8.75 \times 10^{-9} M_\odot$/yr (bottom). The black lines encircle the regions of outward migration. The vertical red lines indicate the ice line at $170$ K. The white line represents the zero-torque line of the fits presented in Appendix~\ref{ap:model}. As $\dot{M}$ drops, the regions of outward migration shrink so that only planets with lower mass can be saved from inward type-I-migration. Additionally, the orbital distance at which outward migration acts becomes smaller with decreasing $\dot{M}$. This is caused by the shallower gradient in temperature for lower $\dot{M}$ discs.
   \label{fig:Migfit}
   }
\end{figure}

In Fig.~\ref{fig:Migfit} the migration map for the $\dot{M}=7 \times 10^{-8} M_\odot$/yr (top) and $\dot{M}= 8.75 \times 10^{-9} M_\odot$/yr disc are displayed. The migration map for $\dot{M}=3.5 \times 10^{-8} M_\odot$/yr can be found in Fig.~\ref{fig:Migsim}. As the disc evolves and $\dot{M}$ decreases, the area of outward migration shrinks and moves inwards. This is caused by the disc becoming colder, and therefore the region of opacity transition at the ice line moves inwards as well. This implies that the strong radial negative gradients in temperature also move inwards, shifting the regions of outward migration to smaller radii (Paper II). In the late stages of the disc evolution, not only is the region of outward migration shifted towards the inner regions of the disc, but it also seems that outward migration is only supported for lower mass planets than for earlier disc evolution stages (high $\dot{M}$). This was already observed in Paper II.

\section{Influence of metallicity}
\label{sec:metallicity}

As the disc evolves in time, the micro-metre dust grains can grow and form larger particles, which (above mm size) do not contribute to the opacity of the disc any more and the opacity of the disc decreases, as the number of micro-metre dust grains reduces. In the later stages of the disc, the number of small dust grains can be replenished because of destructive collisions between planetesimals and larger objects, which would increase the opacity again. We therefore extend our model to range over a variety of metallicities, namely from $0.1\%$ to $3.0\%$.

\begin{figure}[!ht]
 \centering
 \includegraphics[width=0.85\linwx]{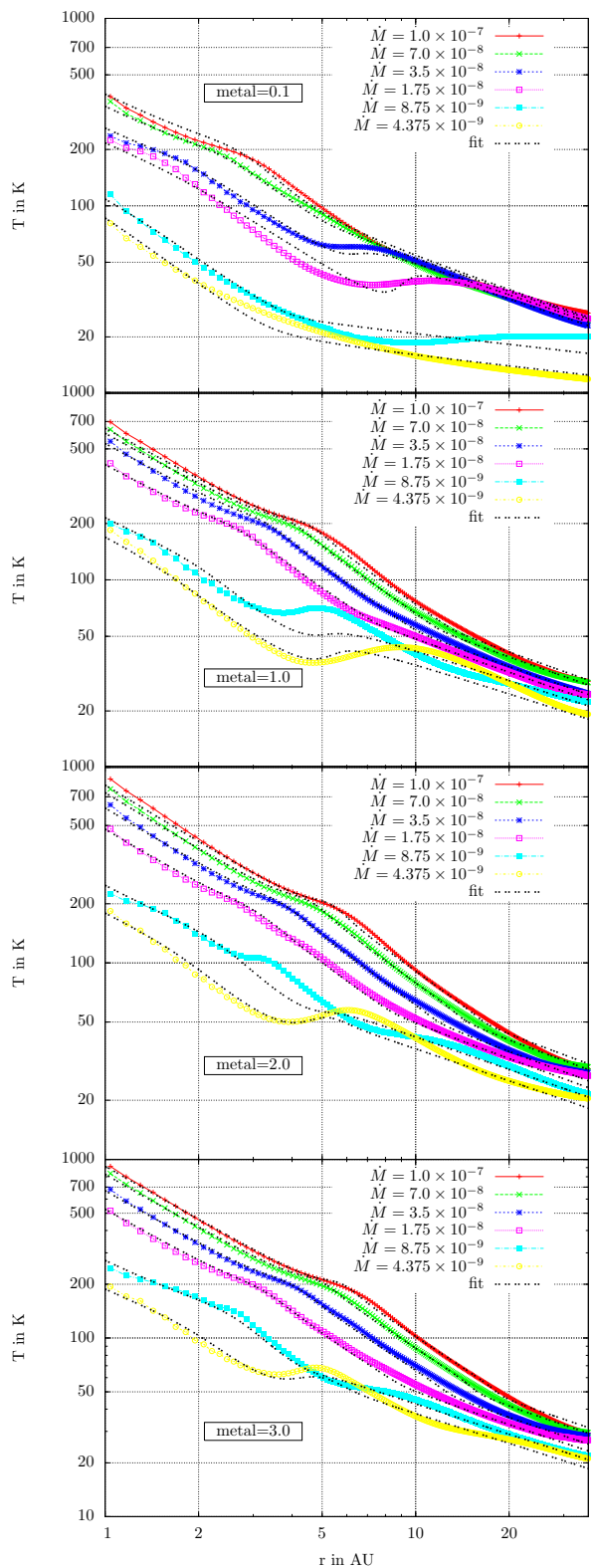}
 \caption{Mid-plane temperature for discs with various $\dot{M}$ and metallicity. The metallicity is $0.1\%$ in the top panel, $1.0\%$ in the second from top panel, $2.0\%$ in the third from top panel, and $3.0\%$ in the bottom panel. An increasing metallicity reduces the cooling rate of the disc, making the disc hotter.
   \label{fig:Tmetalall}
   }
\end{figure}

The temperature for discs with metallicities of $0.1\%$, $1.0\%$, $2.0\%$, and $3.0\%$ (from top to bottom) are displayed in Fig.~\ref{fig:Tmetalall}. The $H/r$ and surface density profiles of those discs are shown in Appendix~\ref{ap:model}. Clearly, discs with less metallicity are colder than their high-metallicity counterparts. This is caused by the increased cooling for low-metallicity discs, because $D \propto 1/(\rho \kappa_R)$. The general trends of the disc structure, however, hold for all metallicities, namely, that all discs feature a bump in temperature at the opacity transition at the ice line. 

However, discs with $\dot{M}<1 \times 10^{-8} M_\odot$/yr and a metallicity of $0.1\%$ no longer show a bump in temperature, and they follow the perfect flaring disc, because they are so cold that the transition in opacity at the ice line is no longer present in the outer parts of the disc. Additionally, the temperature inversion observed in sect.~\ref{sec:timeevolve} occurs at higher $\dot{M}$ rates for low-metallicity discs than for discs with higher metallicity. 

Discs that have the same $\dot{M}$ value, but different metallicity do not have the same surface density profile (see Fig.~\ref{fig:Sigmetalall} in Appendix~\ref{ap:model}). As the opacity increases, the disc becomes hotter because it cools less efficiently, and therefore $H$ increases, which then results in higher viscosity, reducing $\Sigma_G$ as $\dot{M}$ is constant in $r$ (eq.~\ref{eq:Mdothydro}). An increase in metallicity therefore does not result in an increase of the same factor in temperature. The opposite applies for a lower metallicity, which results in a cooler disc (smaller $H$), decreasing the viscosity and therefore increasing $\Sigma_G$.

The increase in temperature is much greater in the inner parts of the disc than in the outer parts. The reason for that lies in the fact that the inner disc is dominated by viscous heating, which does not depend on the opacity, and only the change in the cooling rate depends on the opacity and therefore changes the temperature in the inner disc. In the outer disc, however, the cooling and the heating both depend on the opacity. A reduced opacity increases the cooling, making the disc thinner ($H/r$ decreases), but at the same time, the absorption of stellar photons is also reduced in the upper layers because fewer dust grains are available. This means that stellar irradiation can be efficient at much smaller heights from the mid plane, heating the disc at a smaller vertical height $z$, shifting the heated region closer to the mid plane, and making the disc hotter. This effect counterbalances the increased cooling, so that the disc stays roughly at the same temperature in the parts that are dominated by stellar irradiation for all metallicities, as long as the changes in metallicity are not too big.

The higher temperature in the inner disc, especially for lower $\dot{M}$, caused by higher metallicity has important implications for planetesimal formation, planetary migration, and the evolution of the ice line in time. These are discussed in sect.~\ref{sec:formplanets}.

\section{Discussion}
\label{sec:formplanets}

 Planets will form during the evolution of the protoplanetary accretion disc. We discuss here the general implications of the disc structure on the formation of planetesimals and embryos (sect.~\ref{subsec:SI}) and giant planets (sect.~\ref{subsec:giants}). Additionally, we focus on the inward motion of the ice line as the disc evolves in time (sect.~\ref{subsec:iceline}).

\subsection{Planetesimal and embryo formation}
\label{subsec:SI}

The first building blocks of planetary embryos are planetesimals, which can be formed by the streaming instability \citep{2007ApJ...662..627J}. The streaming instability requires not only a small $\Delta$ parameter (Fig.~\ref{fig:etaall}), but also an increased amount of heavy elements \citep{2010ApJ...722L.220B}. Both of these requirements are satisfied at the ice line, where an increased amount of heavy elements is likely to be present owing to the condensation of ice grains and pebbles \citep{2013A&A...552A.137R}. Additionally this region features a low $\Delta$ value, because it is in the shadowed region of the disc, the minimum of $H/r$. This feature is independent of the overall metallicity of the disc (sect.~\ref{sec:metallicity}), but because the disc is hotter for higher metallicity discs, the $\Delta$ parameter is also larger, as $H/r$ increases. This implies that for higher metallicity discs, the streaming instability might not be able to be operated, because $H/r$ is greater. However, the formation of the first planetesimals and planetary embryos is still most likely at the location of the ice line, since the disc features a minimum of $\Delta$ in that region. In the inner parts of the disc, the $\Delta$ parameter is also smaller. But since the disc is very hot there, the icy particles have evaporated and fewer metal grains are thus available for triggering the streaming instability.

If indeed the first planetesimals form at the ice line, then embryo formation is most likely to occur here as well. These embryos can then grow via collisions between each other and through pebble accretion \citep{2012A&A...544A..32L}. The growth rate versus the migration rate is now key to determining what kind of planet emerges. If the planet grows rapidly to several Earth masses, then it can overcome the inward type-I-migration, be trapped in a region of outward migration \citep{2014arXiv1408.6094L}, and eventually become the core of a giant planet at a large orbital radius. If the planet does not grow fast enough to be trapped in a region of outward migration, the planet migrates inwards and can end up as either a giant planet in the inner system (if the core can grow fast enough in the inner parts of the disc) or `just' be a hot super-Earth or mini-Neptune \citep{2014arXiv1407.6011C}. 

As the disc evolves in time, the $\Delta$ parameter becomes smaller in the inner disc, meaning that the formation of planetesimals might be more likely at later times. However, the farther the disc has evolved in time, the less time is left for giant planets to form, because those have to accrete gas from the surrounding disc. Planetesimal formation might also be triggered indirectly by grain growth. As the metallicity of $\mu m$ dust grains drops, e.g. due to grain growth, $H/r$ decreases, because of a drop in opacity, which increases the cooling of the disc. This then leads to a lower $\Delta$ value, making it easier for the streaming instability to operate. These two effects would imply that the formation of planetesimals is easier in the late stages of the disc evolution, thus potentially explaining the abundance of small planets compared to gas giants \citep{2013ApJ...766...81F}, which require an earlier core formation in order to accrete an gaseous envelope.

In the outer parts of the disc ($r>10$ AU), the $\Delta$ parameter of the $\dot{M}$ discs, regardless of the discs metallicity, is up to $50\%$ smaller than in the MMSN and CY2010 model. A reduction in $\Delta$ by this factor significantly helps the clumping of particles in the streaming instability that can then lead to the formation of planetesimals, making the formation of planetesimals much more likely in our presented model. This makes it much easier to form Kuiper belt objects and the cores of Neptune and Uranus in
our model. During time, as $\Sigma_G$ reduces, the $\Delta$ parameter even reduces slightly in the outer disc, making the formation of planetesimals more likely in the later stages. Such late formation of planetesimals in the outer disc could explain why Neptune and Uranus did not grow to become gas giants \citep{2014arXiv1408.6087L, 2014arXiv1408.6094L}.

A full analysis of the effect of the disc structure on planetary formation and migration will require introducing detailed prescriptions for planetary growth. This will be the topic of a future paper.

\subsection{Giant planet formation}
\label{subsec:giants}

For solar-like stars, the occurrence of a giant planet ($M_{\rm P} > 0.1M_{\rm Jup}$) is related to the metallicity of the host star. In particular, higher metallicity implies a higher occurrence rate of giant planets \citep{2004A&A...415.1153S, 2005ApJ...622.1102F}. This implies that the formation of planets in systems with higher metallicity might be systematically different from systems with lower metallicity. 

In the core accretion scenario, a core forms first, which then accretes gas to become a giant planet. The core itself migrates through the disc as it grows. \citet{2014arXiv1407.6011C} show that planets that are not trapped inside regions of outward migration are more likely to become super-Earths rather than giant planets, making zones of outward migration important for the final structure of planetary systems.

\begin{figure}
 \centering
 \includegraphics[scale=0.71]{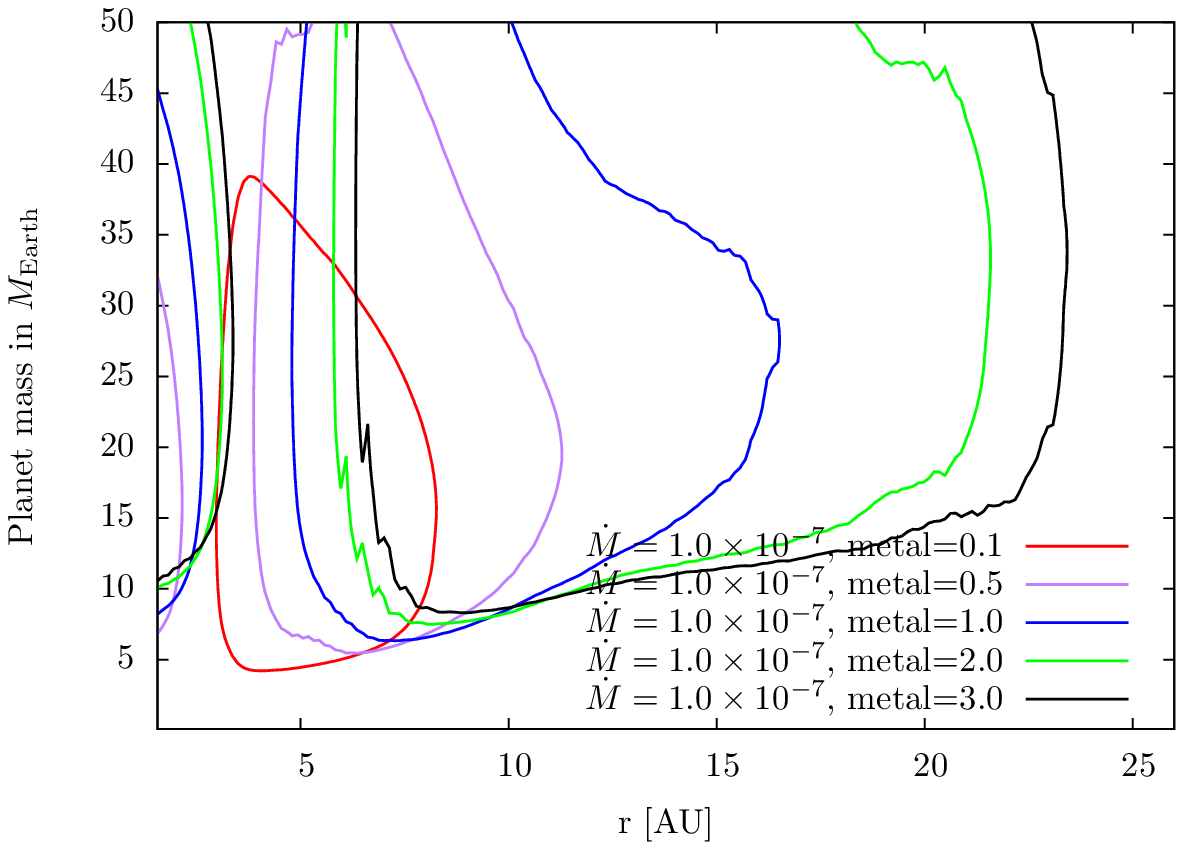}
 \includegraphics[scale=0.71]{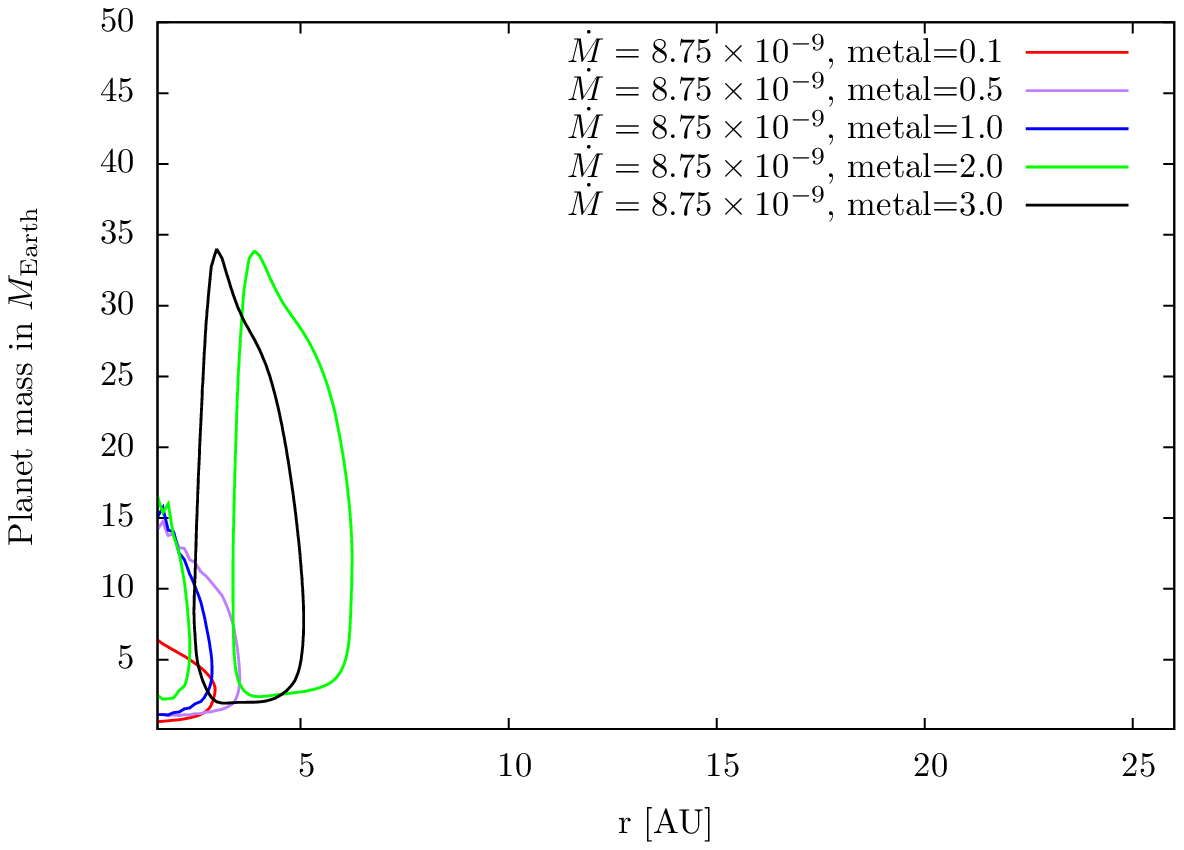}
 \caption{Migration contours for discs with different metallicities. The top panel shows $\dot{M}= 1.0 \times 10^{-7} M_\odot$/yr; the bottom panel shows $\dot{M}= 8.75 \times 10^{-9} M_\odot$/yr. Planets migrate outwards in the regions of the disc that are enclosed by the solid lines, where the colours mark different metallicities.
   \label{fig:Migmetal}
   }
\end{figure}

The mid-plane temperature profiles for discs with different $\dot{M}$ and metallicity are shown in Fig.~\ref{fig:Tmetalall}. They imply that higher metallicity results in a hotter disc that can keep the shadowed regions longer (see also Fig.~\ref{fig:Hrmetalall}), which are able to support outward migration at a few AU for a longer time compared to lower metallicity discs. This is illustrated in Fig.~\ref{fig:Migmetal}, which shows the migration regions where outward migration is possible for discs with different metallicity and for two $\dot{M}$ values; $\dot{M}= 1.0 \times 10^{-7} M_\odot$/yr and $\dot{M}= 8.75 \times 10^{-9} M_\odot$/yr.

For the $\dot{M}= 1.0 \times 10^{-7} M_\odot$/yr discs, there are three clear trends visible for higher metallicities:
\begin{itemize}
 \item the regions of outward migration are shifted farther away from the central star, because the disc is hotter and therefore the opacity transition at the ice line is farther away;
 \item the regions of outward migration are larger in radial extent, because the shadowed regions in the disc are larger, which enlarges the region where a steep radial temperature gradient exits, in turn enlarging the entropy driven corotation torque;
 \item outward migration is only possible for higher minimal masses, which is caused by the changes in the disc structure that overcompensate for the reduced cooling time caused by the higher metallicity, which would actually reduce the minimal mass required for outward migration
\end{itemize}
In principle one could also infer from Fig.~\ref{fig:Migmetal} that outward migration is also possible for higher planetary masses for increasing metallicity. However, as the torques acting on the planets are calculated by a torque formula that was derived in the linear regime for low mass planets ($\approx 5M_{\rm Earth}$), one cannot extend it towards masses of a few $10$ Earth masses. Additionally planets that are that massive start to open up gaps in the disc, which means that the planet transitions into type-II-migration.

The regions of outward migration then shrink and move closer towards the star as the disc evolves in time and $\dot{M}$ reduces. For the $\dot{M}= 8.75 \times 10^{-9} M_\odot$/yr disc, the discs with high metallicity ($>2.0\%$) maintain two large regions of outward migration at a few AU, which are able to trap planets of up to $\approx 30 M_{\rm Earth}$. The lower metallicity discs, on the other hand, only feature one region of outward migration in the inner region of the disc, which is only able to trap cores up to $\approx 15M_{\rm Earth}$.

Additionally, as the regions of outward migration last longer in the outer disc for high metallicity, the growing core can stay farther out, being released at a larger orbital distance into type-II-migration when it becomes large enough. This core would then migrate inwards in the type-II regime and finally be stopped when the disc dissipates \citep{2012MNRAS.422L..82A}, which might explain the pile-up of Jupiter-sized planets around $1$AU. On the other hand, in a low metallicity disc, the planetary core would be trapped closer to the star, and when it is released in type-II-migration as a gas giant it is already closer to the star and might therefore become a hot Jupiter.

\subsection{Ice line}
\label{subsec:iceline}

As the disc evolves in time and $\dot{M}$ drops, the inner regions of the disc become colder, moving the ice line closer to the star. In the evolution of all our disc models, the ice line (at $170$ K) moves to $1$AU at $\approx 2$ Myr. However, this result is troublesome considering evidence from meteors and asteroids in our own solar system. Ordinary and enstatite chondrites contain very little water and must have formed on the warm side of the ice line. The parent bodies formed $2-3$ Myr after CAIs (which mark the zero age of the solar system). However, at this time the ice line in our nominal models had moved to approximately $1$ AU. This is potentially in conflict with the dominance of S-type asteroids, believed to be the source of ordinary chondrites in the inner regions of the asteroid belt. We propose two different ideas that could potentially add to the solution of this problem.

{\it The metallicity} is the key parameter for the temperature profile of the disc. Our simulations show that a higher metallicity in $\mu m$ dust increases the temperature of the inner disc. If the metallicity is even higher (e.g. $5\%$) the ice line would have been farther outside. Therefore an intrinsic higher metallicity of the disc could help this problem. Alternatively, a higher metallicity in $\mu m$-sized dust grains can be created by dust-polluted ice balls that drift across the ice line and then release their dust grains as the ice melts, increasing the metallicity in $\mu m$-sized dust in the inner parts of the disc \citep{2011ApJ...733L..41S}.

{\it The time evolution} of the accretion disc contains large error bars in time \citep{1998ApJ...495..385H}, which could simply imply that the solar system was one of the slower evolving discs, keeping a higher $\dot{M}$ rate at later times, which would keep the ice line farther out and thus the inner system dry at later stages of the disc evolution. For example, in \citet{1998ApJ...495..385H} one observed disc still has $\dot{M}\approx 2.0 \times 10^{-8} M_\odot$/yr at $3$ Myr, which has a high enough temperature to keep the ice line at $\approx 2$ AU for $Z \geq 0.005$. \citet{arXiv:1406.0722} report discs with high accretion rates ($\dot{M}=1 \times 10^{-8} M_\odot$/yr) that are $10$ Myr old, which would then have an ice line farther out and thus avoid the mentioned problem.

However, it is unlikely that these simple scenarios are solely responsible for keeping the snow line far out during the lifetime of the disc, and more complicated scenarios can play a role. This includes local heating from shocks in disc with dead zones \citep{2013MNRAS.434..633M} or the interplay among the radial motion of the gas, drifting icy particles,  and growing planets, which will be the subject of a future paper.

\section{Summary}
\label{sec:summary}

In this work we have compared the power law assumptions of the minimum mass solar nebula and the \citet{2010AREPS..38..493C} models with simulations of protoplanetary discs that feature realistic radiative cooling, viscous, and stellar heating. The modelled disc structures show bumps and dips that are caused by transitions in the opacity, because a change in opacity changes the cooling rate of the disc \citep{2014A&A...564A.135B}. These features can act as sweet spots for forming planetesimals via the streaming instability and for stopping the inward migration of planetary cores of a few Earth masses. Regions of low pressure support also enhance the growth of planetary cores via pebble accretion. These attributes are lacking in the MMSN and CY2010 models, making the formation of planets in these models much harder. Additionally, as the disc evolves in time and the accretion rate $\dot{M}$ decreases, the radial gradients of $\Sigma_G$, $H/r$ and $T$ in the disc change. This temporal evolution is not taken into account in the MMSN and CY2010 models either. 

During the time span of a few Myr, the star evolves along with the disc. The star changes its size and temperature and therefore its luminosity. This changes the amount of stellar heating received by the disc and strongly changes the parts of the disc where stellar heating is the dominant heat source. For higher $\dot{M}$ this mostly affects the outer parts of the disc, while for lower $\dot{M}$, as viscous heating becomes less and less important, the whole disc structure is affected. 

We present a simple fit of our simulated discs over all accretion rates and for different metallicities. The fit consists of three parts that correspond to three different regions in the disc, which are dominated by different heat sources. The inner disc is dominated by viscous heating, while the outer disc is dominated by stellar irradiation. In between, a transition region in the disc exists where viscous heating starts to become less important, but is in the shadow of direct stellar illumination at the same time. 

The different $\dot{M}$ values can be linked to a time evolution of the disc obtained from observations \citep{1998ApJ...495..385H}. With this simple relation between time and accretion rate, our presented fit can easily be used to calculate the disc structure at any given evolution time of the star-disc system. This can then be used as input for planet formation models.

A Fortran script producing $T$, $\Sigma$, and $H/r$ as functions of $Z$, $\dot{M,}$ and time is available upon request.

\begin{acknowledgements}

B.B.,\,A.J.,\,and M.L.\,thank the Knut and Alice Wallenberg Foundation for their financial support. A.J.\,was also
supported by the Swedish Research Council (grant 2010-3710) and the European
Research Council (ERC Starting Grant 278675-PEBBLE2PLANET). A.M.\, is thankful to the Agence Nationale pour la Recherche under grant ANR-13-BS05-0003-01 (MOJO). The computations were done on the “Mesocentre SIGAMM” machine, hosted by the Observatoire de la C\^{o}te d'Azur. We thank M. Havel for discussing with us on the early evolution model of the Sun.

\end{acknowledgements}

\appendix
\section{A simple model for evolving accretion discs}
\label{ap:model}

In this appendix we provide a fit for our simulated disc models for all the different $\dot{M}$ stages. We provide here only a fit for the temperature $T$, because for discs with constant $\dot{M}$, all other quantities can be derived, from the value of $\dot{M}$ and the $\alpha$ parameter. This is done by using the $\dot{M}$ relation and the hydrostatic equilibrium equation
\begin{equation}
\label{eq:Mdothydro}
 \dot{M} = 3 \pi \nu \Sigma_G = 3 \pi \alpha H^2 \Omega_K \Sigma_G \quad , \quad  T = \left( \frac{H}{r} \right)^2 \frac{G M_\star}{r} \frac{\mu}{\cal R} \ , 
\end{equation}
where $\cal R$ is the gas constant. The temperature given in the following fit is the mid-plane temperature of the disc, which does not represent the vertical structure of the disc that can feature a super-heated upper layer (e.g. Paper I). With the fit of the temperature, $H$ can be determined and then finally $\Sigma_G$. 

\begin{figure}
 \centering
 \includegraphics[width=1.0\linwx]{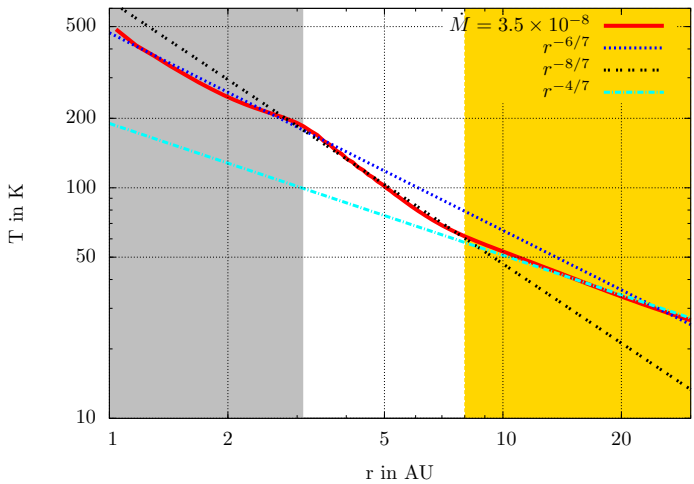}
 \caption{Radial temperature profile of the disc with $\dot{M}=3.5 \times 10^{-8} M_\odot$/yr. The different background colours mark different fit-regimes. The grey area marks the inner part of the disc that is dominated by viscous heating, while the golden part indicates the outer parts of the disc that is dominated by stellar heating. The white middle region indicates the shadowed region of the disc.
   \label{fig:Tempregion}
   }
\end{figure}

We identify three different regimes for the temperature gradients in the disc. The inner disc is dominated by viscous heating and features a
higher temperature than the water condensation temperature (Fig.~\ref{fig:Tempregion}), while the outer disc is dominated by stellar irradiation. In between, the disc is in the shadow of stellar irradiation. This can be seen more clearly in the middle panel of Fig.~\ref{fig:HrTSigallfit}, where $H/r$ drops with increasing $r$. We fit each of these three regimes with individual power laws, which are indicated in Fig.~\ref{fig:Tempregion} as well.

The fit for the temperature for the outer part of the disc is $T\propto r^{-4/7}$ which is steeper than in the MMSN and CY2010 models that are based on analytical arguments (e.g. \citet{1997ApJ...490..368C}). The reason for this originates in the shadowed parts of the disc. The disc is cooler in the shadow and is heated by diffusion from the hotter inner part (heated viscously) and by the hotter outer part (heated from the star). This leads to a steeper temperature gradient in the outer parts, where the disc just emerges from the shadow, since it has to (partially) compensate for the lack of heating in the shadowed region. This compensation for heat influences the gradient in temperature to quite large distances. But we suspect that the gradient in temperature will return back to $T\propto r^{-3/7}$ in the very outer parts of the disc, because the influence of the shadowed region reduces with larger orbital distances. This is also true for discs in the later stage evolution, since those discs no longer feature a prominent bump at the ice line that can shade parts of the disc from stellar irradiation.

The changes in the temperature gradient originate in the changes of opacity, which changes the cooling rate of the disc and hence the temperature (Paper II). The changes in opacity occur at the ice line, where ice grains melt and sublimate, and at the silicate line ($\approx 1000$K). As $\dot{M}$ decreases, the viscous heating in the inner disc decreases as well, making the disc colder and moving the ice-line bump closer towards the central star. If the temperature of the disc is always below the water condensation temperature, the disc will therefore not feature the ice line bump any more (see $\dot{M}<8.75 \times 10^{-8} M_\odot$/yr in Fig.~\ref{fig:HrTSigallfit}), and thus the inner part of the fit, which requires a higher temperature than the water condensation temperature, is not needed any more. Consequently, the fit of the temperature profile will only consist of fits of the middle and outer parts of the disc (indicated in white and yellow in Fig.~\ref{fig:Tempregion}).

As long as the temperature is below the sublimation temperature of silicate grains ($T<1000$K), the fit can be applied to regions that are inside $1$AU. The reason for that is that there are no additional bumps in the opacity profile for $170$K$<T<1000$K that could change the disc structure by changing the cooling rate of the disc (Paper I).

As $\dot{M}$ decreases and the viscous heating in the inner parts of the disc decreases, the disc's heating is dominated more and more by stellar irradiation. This transition in the dominant heating process occurs for $\dot{M} \approx 1.75 \times 10^{-8} M_\odot$/yr. We therefore observe a huge drop in the disc's temperature (top panel in Figs.~\ref{fig:HrTSigallfit} and~\ref{fig:Tmetalall}) around these $\dot{M}$ rates. For even lower $\dot{M}$ rates, the changes in temperature seem to be smaller again, which is caused by the fact that the disc's heating is now entirely dominated by stellar irradiation. We therefore introduce different fitting regimes for different $\dot{M}$ values.

Since the heating and cooling of the disc is directly proportional to the $\dot{M}$ value, our fit is a function of $\dot{M}$. We express each of the three individual power law fits in the following form
\begin{equation}
 f(r) = A B^{\mu_R} r^{-l} \psi_{ij} \ ,
\end{equation}
where $A$ is a reference temperature expressed in K and $B$ the factor by which the scaling with $\dot{M}$ is expressed through $\mu_R = \log (\dot{M}/\dot{M}_{\text{ref}})$, where $\dot{M}_{\text{ref}}$ is the reference $\dot{M}$ value. The factor $\psi_{ij}$ expresses the transition from regime $i$ to regime $j$. The factor $l$ corresponds to the power law index of the fit, shown in Fig.~\ref{fig:Tempregion}. 

In the inner regions of the disc for high $\dot{M}$, the disc is dominated by viscous heating, and higher metallicity reduces the cooling, making the disc hotter, therefore increasing the temperature and the gradient of the temperature, which is then a function of metallicity. For $\dot{M} < 1.75 \times 10^{-8} M_\odot$/yr, the disc starts to become dominated by stellar irradiation, so that the temperature gradient in the inner parts of the disc flattens. At the same $\dot{M}$ rates, the shadowed regions of the disc shrink, making the gradient of temperature in the outer parts of the disc flatter as well, because the outer parts of the disc no longer have to compensate for such a large unheated shadowed region.

The fit for the whole disc with $\dot{M} \geq 1.75 \times 10^{-8} M_\odot$/yr can therefore be expressed as
\begin{eqnarray}
\label{eq:Tfitting}
 T(r,\mu_R,Z) &=& A_1 B_1^{\mu_R} \left( \frac{r}{\rm AU} \right)^{-6/7-\log(100Z+0.5)/4} \psi_{12} \nonumber \\
 &+& A_2 B_2^{\mu_R} \left( \frac{r}{\rm AU} \right)^{-8/7} \psi_{21} \psi_{23} \nonumber \\
 &+& A_3 B_3^{\mu_R} \left( \frac{r}{\rm AU} \right)^{-4/7} \psi_{32} \ .\end{eqnarray}
For $\dot{M} < 1.75 \times 10^{-8} M_\odot$/yr and $\dot{M} \geq 8.75 \times 10^{-9} M_\odot$/yr, the fit is expressed as 
\begin{eqnarray}
\label{eq:Tfitting2}
 T(r,\mu_R,Z) &=& A_1 B_1^{\mu_R} \left( \frac{r}{\rm AU} \right)^{-6/7-\log(100Z+0.5)/4 \times (1-\mu_R/\log(0.5))} \psi_{12} \nonumber \\
 &+& A_2 B_2^{\mu_R} \left( \frac{r}{\rm AU} \right)^{-8/7} \psi_{21} \psi_{23} \nonumber \\
 &+& A_3 B_3^{\mu_R} \left( \frac{r}{\rm AU} \right)^{-4/7 -\mu_R(1/(100Z))/8} \psi_{32} \ ,
\end{eqnarray}
where the gradient of the inner part of the disc relaxes back to $r^{-6/7}$ at $\dot{M} = 8.75 \times 10^{-9} M_\odot$/yr, and the gradient of the outer part of the disc becomes less steep as a function of $Z$. For $\dot{M} < 8.75 \times 10^{-9} M_\odot$/yr the fit is expressed as 
\begin{eqnarray}
\label{eq:Tfitting3}
 T(r,\mu_R,Z) &=& A_1 B_1^{\mu_R} \left( \frac{r}{\rm AU} \right)^{-6/7} \psi_{12} \nonumber \\
 &+& A_2 B_2^{\mu_R} \left( \frac{r}{\rm AU} \right)^{-8/7} \psi_{21} \psi_{23} \nonumber \\
 &+& A_3 B_3^{\mu_R} \left( \frac{r}{\rm AU} \right)^{-4/7 -\log(0.5) \times (1/(100Z))/8} \psi_{32} \ .
\end{eqnarray}
These formula are used for both high and low metallicity discs.

The middle part contains two $\psi$ functions since the second regime has to be smoothed into the first and into the third regime. For all different $\dot{M}$ fitting regimes the values of $A$, $B$, and $\psi$ change, but keep the same shape.

Along with the changes in $\dot{M}$, the disc changes as different metallicities provide different cooling rates and therefore different temperatures. These changes in temperature then reflect back on changes in $H/r$ and $\Sigma_G$. We therefore also introduce different regimes for the metallicity, where we distinguish between high metallicity ($Z \geq 0.005$) and low metallicity ($Z < 0.005$). The factors $A$, $B$, and $\psi$  are therefore functions of $\dot{M}$ and $Z$. The scaling with $Z$ is expressed through $\chi = \log (100Z/0.5)/ \log(2)$.

\subsection{Low metallicity discs}
\label{subsec:lowmet}

In this section the fit for discs with $Z < 0.005$ is presented. The first regime of $\dot{M}$ is for $\dot{M} > 3.5 \times 10^{-8} M_\odot$/yr, which corresponds to the early evolution of the disc. We define $\dot{M}_{\text{ref}} = 3.5 \times 10^{-8} M_\odot$/yr, and the fit reads as (where $r$ is in units of AU)
\begin{eqnarray}
 \label{eq:Thigh}
 A_1 &=& 410 \times 1.4^\chi {\rm K} \nonumber \\
 B_1 &=& 1.3 \times 0.7^\chi \nonumber \\
 \psi_{12} &=& - \big[ \arctan ( r- 3.0 ( 1.832^{\mu_R} + (0.15+0.05\mu_R^2) \chi)) \nonumber \\ &-& \pi/2 \big] /\pi \nonumber \\
 A_2 &=& 650 \times 1.3^\chi {\rm K} \nonumber \\
 B_2 &=& 1.8 \times 0.75^\chi \nonumber \\
 \psi_{21} &=& \big[ \arctan ( r- 2.6 (1.832^{\mu_R}+(0.1666+0.05\mu_R^2) \chi)) \nonumber \\ &+& \pi/2 \big] /\pi \nonumber \\
 \psi_{23} &=& \big[ -\arctan ( r- 8.3 ( 1.372^{\mu_R} + (0.10+\mu_R^2) \chi)) \nonumber \\ &+& \pi/2 \big]/\pi \nonumber \\
 A_3 &=& 185 \times 0.995^\chi {\rm K} \nonumber \\
 B_3 &=& 1.4 \times 1.1^\chi \nonumber \\
 \psi_{32} &=& \big[ \arctan ( r- 8.3 (1.372^{\mu_R} +(0.10+\mu_R^2) \chi )) \nonumber \\ &+& \pi/2 \big] /\pi \ .
\end{eqnarray}
The next regime spans the interval $1.75 \times 10^{-8} M_\odot/\text{yr} < \dot{M} \leq 3.5 \times 10^{-8} M_\odot$/yr and $\dot{M}_{\text{ref}} = 3.5 \times 10^{-8} M_\odot$/yr, and the fit reads
as\begin{eqnarray}
 \label{eq:Tmiddle1}
 A_1 &=& 410 \times 1.4^\chi {\rm K} \nonumber \\
 B_1 &=& 3.0 \times 0.8^\chi \nonumber \\
 \psi_{12} &=& - \big[ \arctan ( r- 3.0 ( 1.832^{\mu_R} + (0.15+0.2\mu_R^2) \chi)) \nonumber \\ &-& \pi/2 \big] /\pi \nonumber \\
 A_2 &=& 650 \times 1.3^\chi {\rm K} \nonumber \\
 B_2 &=& 2.1 \times 1.1^\chi \nonumber \\
 \psi_{21} &=& \big[ \arctan ( r- 2.6 (1.832^{\mu_R}+(0.1666+0.2\mu_R^2) \chi)) \nonumber \\ &+& \pi/2 \big] /\pi \nonumber \\
 \psi_{23} &=& \big[ -\arctan ( r- 8.3 ( 1.372^{\mu_R} + (0.10- 2.5\mu_R^2) \chi)) \nonumber \\ &+& \pi/2 \big]/\pi \nonumber \\
 A_3 &=& 185 \times 0.995^\chi {\rm K} \nonumber \\
 B_3 &=& 1.176 \times 1.01^\chi \nonumber \\
 \psi_{32} &=& \big[ \arctan ( r- 8.3 (1.372^{\mu_R} +(0.10-2.5\mu_R^2) \chi )) \nonumber \\ &+& \pi/2 \big] /\pi \ .
\end{eqnarray}
The next regime spans the interval $8.75 \times 10^{-9} M_\odot/\text{yr} < \dot{M} \leq 1.75 \times 10^{-8} M_\odot$/yr and marks the regime where $T$ undergoes a large change in the inner parts of the disc, because viscous heating is replaced by stellar heating as the dominant heat source. Here, $\dot{M}_{\text{ref}} = 1.75 \times 10^{-8} M_\odot$/yr (which changes $\mu_R$), which results in some reduction factors $R$ that need to be taken into account:
\begin{eqnarray}
 R_{12} &=& 1.832^{\log(0.5)}+ [0.15+0.2(\log(0.5))^2]\chi \nonumber \\ R_{21} &=& 1.832^{\log(0.5)}+ [0.1666+0.2(\log(0.5))^2]\chi \nonumber \\
 R_{23} &=& 1.372^{\log(0.5)}+ [0.10-2.5(\log(0.5))^2]\chi \nonumber \\ R_{32} &=& 1.372^{\log(0.5)}+ [0.10-2.5(\log(0.5))^2]\chi \ .    
\end{eqnarray}
The fit then reads
as\begin{eqnarray}
\label{eq:Tmiddle2}
 A_1 &=& 410 \times (3\times 0.8^\chi)^{\log(0.5)} \times 1.4^\chi {\rm K} \nonumber \\
 B_1 &=& 8.0 \times 1.5^\chi \nonumber \\
 \psi_{12} &=& -\left[ \arctan ( r- 3.0 R_{12} (3-0.2\chi^2)^{\mu_R}) -\pi/2 \right] /\pi \nonumber \\
 A_2 &=& 650 \times (2.1 \times 1.1^\chi)^{\log(0.5)} \times 1.3^\chi {\rm K} \nonumber \\
 B_2 &=& 30.0 \times 1.1^\chi \nonumber \\
 \psi_{21} &=& \left[ \arctan ( r- 2.6 R_{21} (5-0.2\chi^2)^{\mu_R})+\pi/2 \right] /\pi \nonumber \\
 \psi_{23} &=& \left[ -\arctan ( r- 8.3 R_{23} (3-4.5\chi)^{\mu_R})+\pi/2 \right] /\pi \nonumber \\
 A_3 &=& 185 \times (1.176\times 1.01^\chi)^{\log(0.5)} \times 0.995^\chi {\rm K} \nonumber \\
 B_3 &=& 2.2 \times 0.125^\chi \nonumber \\
 \psi_{32} &=& \left[ \arctan ( r- 8.3 R_{32} (3.5-4.5\chi)^{\mu_R} ) + \pi/2 \right] /\pi \ .
\end{eqnarray}
For even lower $\dot{M}$ values, the disc is dominated by stellar heating. We therefore have to introduce another regime that is $\dot{M} \leq 8.75 \times 10^{-9} M_\odot$/yr. The reduction factors are now
\begin{eqnarray}
 R_{12} &=& 1.832^{\log(0.5)}+ [0.15+0.2(\log(0.5))^2]\chi \nonumber \\ &\times& (3-0.2\chi^2)^{\log(0.5)} \nonumber \\
 R_{21} &=& 1.832^{\log(0.5)}+ [0.1666+0.2(\log(0.5))^2]\chi \nonumber \\ &\times& (5-0.2\chi^2)^{\log(0.5)} \nonumber \\
 R_{23} &=& 1.372^{\log(0.5)}+ [0.10-2.5(\log(0.5))^2]\chi \nonumber \\ &\times& (3-4.5\chi)^{\log(0.5)} \nonumber \\
 R_{32} &=& 1.372^{\log(0.5)}+ [0.10-2.5(\log(0.5))^2]\chi \nonumber \\ &\times& (3.5-4.5\chi)^{\log(0.5)} \ .    
\end{eqnarray}
Here $\dot{M}_{\text{ref}} = 8.75 \times 10^{-9} M_\odot$/yr and the fit reads
as\begin{eqnarray}
\label{eq:Tlow}
 A_1 &=& 410 \times (3\times 0.8^\chi)^{\log(0.5)} \times (8\times 1.5^\chi)^{\log(0.5)} \times 1.4^\chi {\rm K} \nonumber \\
 B_1 &=& 2.0 \times 1.4^\chi \nonumber \\
 \psi_{12} &=& -\left[ \arctan ( r- 3.0 R_{12} \times 3^{\mu_R}) -\pi/2 \right] /\pi \nonumber \\
 A_2 &=& 650 \times (2.1 \times 1.1^\chi)^{\log(0.5)} \times (30.0 \times 1.1^\chi)^{\log(0.5)} \times 1.3^\chi {\rm K} \nonumber \\
 B_2 &=& 1.5 \times 0.7^\chi \nonumber \\
 \psi_{21} &=& \left[ \arctan ( r- 2.6 R_{21} \times 5^{\mu_R})+\pi/2 \right] /\pi \nonumber \\
 \psi_{23} &=& \left[ -\arctan ( r- 8.3 R_{23} (-0.5\chi-2\mu_R)^{\mu_R})+\pi/2 \right] /\pi \nonumber \\
 A_3 &=& 185 \times (1.176\times 1.01^\chi)^{\log(0.5)} \times (2.2\times 0.125^\chi)^{\log(0.5)} \nonumber \\ &\times& 0.995^\chi {\rm K} \nonumber \\
 B_3 &=& 1.9 \times 0.9^\chi \nonumber \\
 \psi_{32} &=& \left[ \arctan ( r- 8.3 R_{32} (-0.5\chi-2\mu_R)^{\mu_R} ) + \pi/2 \right] /\pi \ .
\end{eqnarray}

For accretion rates of $1 \times 10^{-10} M_\odot/\text{yr} < \dot{M} < 1 \times 10^{-8} M_\odot$/yr, photo evaporation of the disc can be dominant and clear the disc on a very short time scale \citep{2013arXiv1311.1819A}. This implies that going to accretion rates of $\dot{M} < 1 \times 10^{-9} M_\odot$/yr might not be relevant for the evolution of the disc. Therefore we recommend not using these fits for $\dot{M} < 1 \times 10^{-9} M_\odot$/yr because the disc will be photo-evaporated away in a very short time for these parameters.

In Fig.~\ref{fig:HrTSigallfit} the mid-plane temperature (top), $H/r$ (middle), and the surface density (bottom) of our simulations and the fits from eq.~\ref{eq:Thigh} to eq.~\ref{eq:Tlow} with $Z=0.005$ are displayed (shown in black lines). The general quantities of the discs are captured well by the provided fits. In Fig.~\ref{fig:etaall} the $\Delta$ parameter (eq.~\ref{eq:stream}) of the simulations with $Z=0.005$ and the one derived from the fit is displayed. The general trend towards a decreasing $\Delta$ in the inner parts of the disc with decreasing $\dot{M}$ is reproduced well. The pressure in the disc is calculated through its proportionality with the disc's density
\begin{equation}
 \label{eq:pressure}
 P_{3D} = \mathcal{R} \rho_G T / \mu = (\gamma - 1) \rho_G c_V T \ ,
\end{equation}
where $c_V$ is the specific heat at constant volume. In hydrostatic equilibrium the density is calculated through $\rho_G = \Sigma_G / (\sqrt{2 \pi} H)$. The fit in $\Delta$ seems to be a bit off for $\dot{M} < 4.375 \times 10^{-9} M_\odot$/yr. The reason for that lies in the different quantities that are needed to compute $\Delta$. Small errors in independent variables (e.g. $T$, $\Sigma_G$) translate into larger errors in the final quantity (here $\Delta$).

In Fig.~\ref{fig:Migfit} the torque maps for the discs with $\dot{M}=7 \times 10^{-8} M_\odot$/yr (top) and $\dot{M}= 8.75 \times 10^{-9} M_\odot$/yr (bottom) with $Z=0.005$ are displayed. The white lines indicate the zero-torque lines resulting from the fitted disc structures presented by the black lines in Fig.~\ref{fig:HrTSigallfit}. The overall agreement in the torque acting on the planet from the fit compared to the simulation is quite good, but there seem to be some variations on the `edges' of the region of outward migration. This is caused by the fit having small deviations from the simulations for all quantities of the disc. But in the migration formula of \citet{2011MNRAS.410..293P}, all these quantities and their gradients play a role. This is especially important for the torque saturation at higher planetary masses. However, the general trend, especially for low mass planets, is captured well.

\subsection{High metallicity discs}
\label{subsec:highmet}

\begin{figure}[!ht]
 \centering
 \includegraphics[width=0.85\linwx]{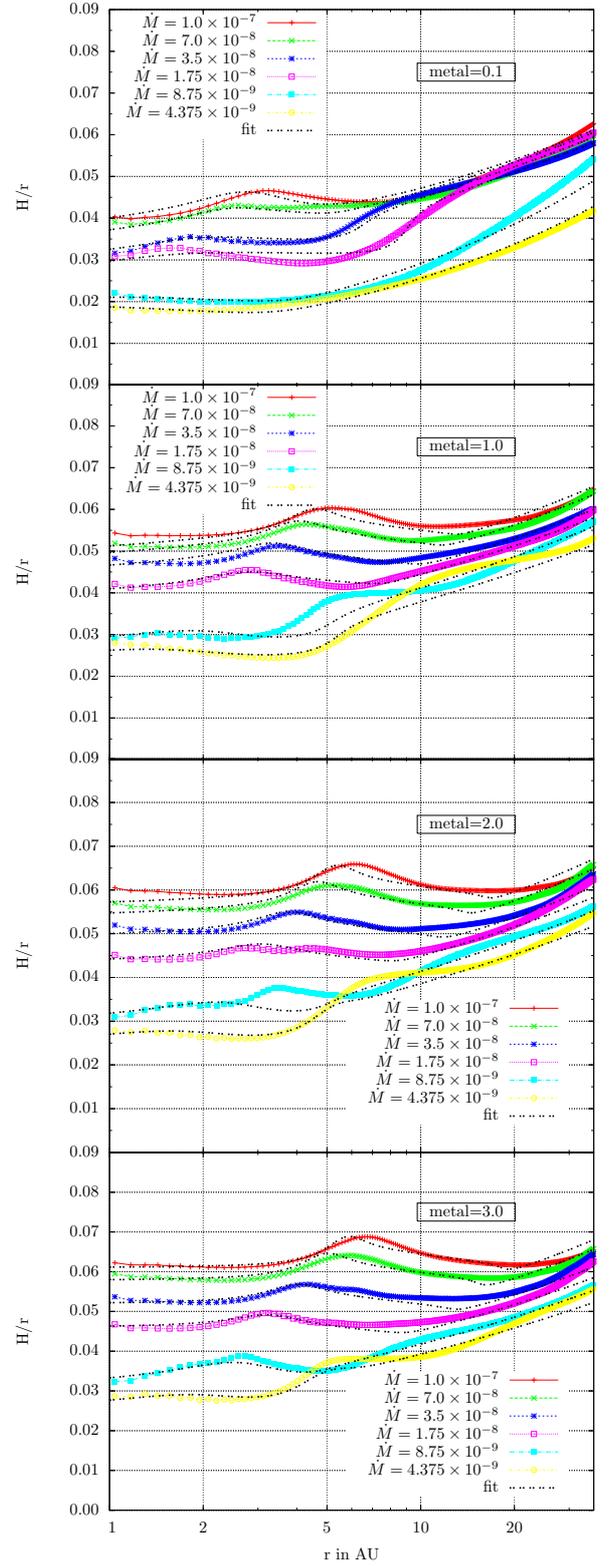}
 \caption{$H/r$ for discs with various $\dot{M}$ and metallicity. The metallicity is $0.1\%$ in the top panel, $1.0\%$ in the second from top panel, $2.0\%$ in the third from top panel, and $3.0\%$ in the bottom panel. An increasing metallicity reduces the cooling rate of the disc, making the disc hotter, and therefore increases $H/r$. The black lines in the plots indicate the fit presented in appendix~\ref{ap:model}.
   \label{fig:Hrmetalall}
   }
\end{figure}

\begin{figure}[!ht]
 \centering
 \includegraphics[width=0.85\linwx]{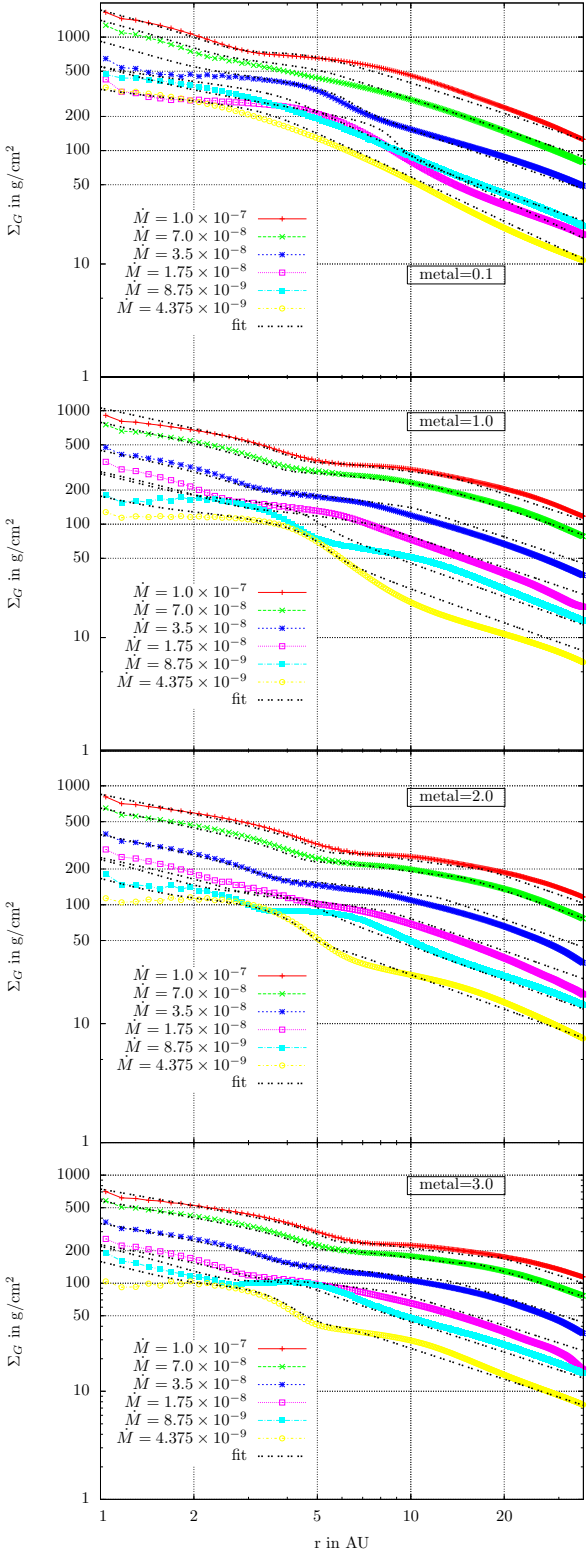}
 \caption{Surface density $\Sigma_G$ for discs with various $\dot{M}$ and metallicity. The metallicity is $0.1\%$ in the top panel, $1.0\%$ in the second from top panel, $2.0\%$ in the third from top panel, and $3.0\%$ in the bottom panel. An increasing metallicity reduces the cooling rate of the disc, making the disc hotter, increasing the viscosity, and therefore decreasing $\Sigma_G$ compared to discs with lower metallicity. The black lines in the plots indicate the fits presented in appendix~\ref{ap:model}.   \label{fig:Sigmetalall}
   }
\end{figure}

In this section fits for discs with high metallicity ($Z \geq 0.005$) are presented. The fits follow the basic principles as for the low metallicity discs, but the parameters for $A$, $B$, and $\psi$ differ, because the dependence on $Z$ and therefore on $\chi$ changes. The first regime of $\dot{M}$ for $\dot{M} > 3.5 \times 10^{-8} M_\odot$/yr is represented by the following fit (where $r$ is in units of AU)
\begin{eqnarray}
 \label{eq:Thighhm}
 A_1 &=& 410 \times 1.175^\chi {\rm K} \nonumber \\
 B_1 &=& 1.3 \times 1.2^\chi \nonumber \\
 \psi_{12} &=& - \big[ \arctan ( r- 3.0 ( 1.832^{\mu_R} + (0.15+0.55\mu_R^2) \chi)) \nonumber \\ &-& \pi/2 \big] /\pi \nonumber \\
 A_2 &=& 650 \times 1.15^\chi {\rm K} \nonumber \\
 B_2 &=& 1.8 \times 1.15^\chi \nonumber \\
 \psi_{21} &=& \big[ \arctan ( r- 2.6 (1.832^{\mu_R}+(0.1666+0.666\mu_R^2) \chi)) \nonumber \\ &+& \pi/2 \big] /\pi \nonumber \\
 \psi_{23} &=& \big[ -\arctan ( r- 8.3 ( 1.372^{\mu_R} + (0.20+\mu_R^2) \chi)) \nonumber \\ &+& \pi/2 \big]/\pi \nonumber \\
 A_3 &=& 185 \times 1.025^\chi {\rm K} \nonumber \\
 B_3 &=& 1.4 \times 1.05^\chi \nonumber \\
 \psi_{32} &=& \big[ \arctan ( r- 8.3 (1.372^{\mu_R} +(0.20+\mu_R^2) \chi )) \nonumber \\ &+& \pi/2 \big] /\pi \ .
\end{eqnarray}
The next regime spans the interval $1.75 \times 10^{-8} M_\odot/\text{yr} < \dot{M} \leq 3.5 \times 10^{-8} M_\odot$/yr and $\dot{M}_{\text{ref}} = 3.5 \times 10^{-8} M_\odot$/yr and the fit reads
as\begin{eqnarray}
 \label{eq:Tmiddle1hm}
 A_1 &=& 410 \times 1.175^\chi {\rm K} \nonumber \\
 B_1 &=& 3.0 \times 0.9^\chi \nonumber \\
 \psi_{12} &=& - \big[ \arctan ( r- 3.0 ( 1.832^{\mu_R} + (0.15-0.4\mu_R^2) \chi)) \nonumber \\ &-& \pi/2 \big] /\pi \nonumber \\
 A_2 &=& 650 \times 1.15^\chi {\rm K} \nonumber \\
 B_2 &=& 2.1 \times 1.15^\chi \nonumber \\
 \psi_{21} &=& \big[ \arctan ( r- 2.6 (1.832^{\mu_R}+(0.1666-0.5\mu_R^2) \chi)) \nonumber \\ &+& \pi/2 \big] /\pi \nonumber \\
 \psi_{23} &=& \big[ -\arctan ( r- 8.3 ( 1.372^{\mu_R} + (0.20- 1.5\mu_R^2) \chi)) \nonumber \\ &+& \pi/2 \big]/\pi \nonumber \\
 A_3 &=& 185 \times 1.025^\chi {\rm K} \nonumber \\
 B_3 &=& 1.176 \times 1.05^\chi \nonumber \\
 \psi_{32} &=& \big[ \arctan ( r- 8.3 (1.372^{\mu_R} +(0.20-1.5\mu_R^2) \chi )) \nonumber \\ &+& \pi/2 \big] /\pi \ .
\end{eqnarray}
The next regime spans the interval $8.75 \times 10^{-9} M_\odot/\text{yr} < \dot{M} \leq 1.75 \times 10^{-8} M_\odot$/yr and marks the regime where $T$ undergoes a large change in the inner parts of the disc, because viscous heating is replaced by stellar heating as dominant heat source. Here, $\dot{M}_{\text{ref}} = 1.75 \times 10^{-8} M_\odot$/yr (which changes $\mu_R$), which results in some reduction factors $R$ that need to be taken into account:
\begin{eqnarray}
 R_{12} &=& 1.832^{\log(0.5)}+ [0.15-0.4(\log(0.5))^2]\chi \nonumber \\ R_{21} &=& 1.832^{\log(0.5)}+ [0.1666-0.5(\log(0.5))^2]\chi \nonumber \\
 R_{23} &=& 1.372^{\log(0.5)}+ [0.2-1.5(\log(0.5))^2]\chi \nonumber \\
 R_{32} &=& 1.372^{\log(0.5)}+ [0.2-1.5(\log(0.5))^2]\chi \ .    
\end{eqnarray}
The fit then reads
as\begin{eqnarray}
\label{eq:Tmiddle2hm}
 A_1 &=& 410 \times (3\times 0.9^\chi)^{\log(0.5)} \times 1.175^\chi {\rm K} \nonumber \\
 B_1 &=& 8.0 \times 1.2^\chi \nonumber \\
 \psi_{12} &=& -\left[ \arctan ( r- 3.0 R_{12} (3-0.2\chi^2)^{\mu_R}) -\pi/2 \right] /\pi \nonumber \\
 A_2 &=& 650 \times (2.1 \times 1.15^\chi)^{\log(0.5)} \times 1.15^\chi {\rm K} \nonumber \\
 B_2 &=& 30.0 \times (1.2-\chi/(\chi+3))^\chi \nonumber \\
 \psi_{21} &=& \left[ \arctan ( r- 2.6 R_{21} (5-0.2\chi^2)^{\mu_R})+\pi/2 \right] /\pi \nonumber \\
 \psi_{23} &=& \left[ -\arctan ( r- 8.3 R_{23} (3+\chi)^{\mu_R})+\pi/2 \right] /\pi \nonumber \\
 A_3 &=& 185 \times (1.176\times 1.05^\chi)^{\log(0.5)} \times 1.025^\chi {\rm K} \nonumber \\
 B_3 &=& 2.2 \times 0.9^\chi \nonumber \\
 \psi_{32} &=& \left[ \arctan ( r- 8.3 R_{32} (3.5+\chi)^{\mu_R} ) + \pi/2 \right] /\pi \ .
\end{eqnarray}
For even lower $\dot{M}$ values, the disc is dominated by stellar heating. We therefore have to introduce another regime that is $\dot{M} \leq 8.75 \times 10^{-9} M_\odot$/yr. The reduction factors are now
\begin{eqnarray}
 R_{12} &=& 1.832^{\log(0.5)}+ [0.15-0.4(\log(0.5))^2]\chi \nonumber \\ &\times& (3-0.2\chi^2)^{\log(0.5)} \nonumber \\
 R_{21} &=& 1.832^{\log(0.5)}+ [0.1666-0.5(\log(0.5))^2]\chi \nonumber \\ &\times& (5-0.2\chi^2)^{\log(0.5)} \nonumber \\
 R_{23} &=& 1.372^{\log(0.5)}+ [0.2-1.5(\log(0.5))^2]\chi \nonumber \\ &\times& (3+\chi)^{\log(0.5)} \nonumber \\
 R_{32} &=& 1.372^{\log(0.5)}+ [0.2-1.5(\log(0.5))^2]\chi \nonumber \\ &\times& (3.5+\chi)^{\log(0.5)} \ .    
\end{eqnarray}
Here $\dot{M}_{\text{ref}} = 8.75 \times 10^{-9} M_\odot$/yr and the fit reads as
\begin{eqnarray}
\label{eq:Tlowhm}
 A_1 &=& 410 \times (3\times 0.9^\chi)^{\log(0.5)} \times (8\times 1.2^\chi)^{\log(0.5)} \times 1.175^\chi {\rm K} \nonumber \\
 B_1 &=& 2.0 \times 1.4^\chi \nonumber \\
 \psi_{12} &=& -\left[ \arctan ( r- 3.0 R_{12} \times 3^{\mu_R}) -\pi/2 \right] /\pi \nonumber \\
 A_2 &=& 650 \times (2.1 \times 1.15^\chi)^{\log(0.5)} \nonumber \\ &\times& [30.0 \times (1.2-\chi/(\chi+3))^\chi]^{\log(0.5)} \times 1.15^\chi {\rm K} \nonumber \\
 B_2 &=& 1.5 \times 1.8^\chi \nonumber \\
 \psi_{21} &=& \left[ \arctan ( r- 2.6 R_{21} \times 5^{\mu_R})+\pi/2 \right] /\pi \nonumber \\
 \psi_{23} &=& \left[ -\arctan ( r- 8.3 R_{23} (0.1\chi-2\mu_R)^{\mu_R})+\pi/2 \right] /\pi \nonumber \\
 A_3 &=& 185 \times (1.176\times 1.05^\chi)^{\log(0.5)} \times (2.2\times 0.9^\chi)^{\log(0.5)} \nonumber \\ &\times& 1.025^\chi {\rm K} \nonumber \\
 B_3 &=& 1.9 \times 0.9^\chi \nonumber \\
 \psi_{32} &=& \left[ \arctan ( r- 8.3 R_{32} (0.2\chi^2-2\mu_R)^{\mu_R} ) + \pi/2 \right] /\pi \ .
\end{eqnarray}
The black lines in Fig.~\ref{fig:Tmetalall} represent the fits obtained with the formulas of eq.~\ref{eq:Thighhm} to eq.~\ref{eq:Tlowhm} for discs with different $Z$. In Figs.~\ref{fig:Hrmetalall} and~\ref{fig:Sigmetalall}, the surface density and $H/r$ of the same discs and the corresponding fits are displayed.

\subsection{Time evolution of the disc}

\begin{figure}
 \centering
 \includegraphics[scale=0.71]{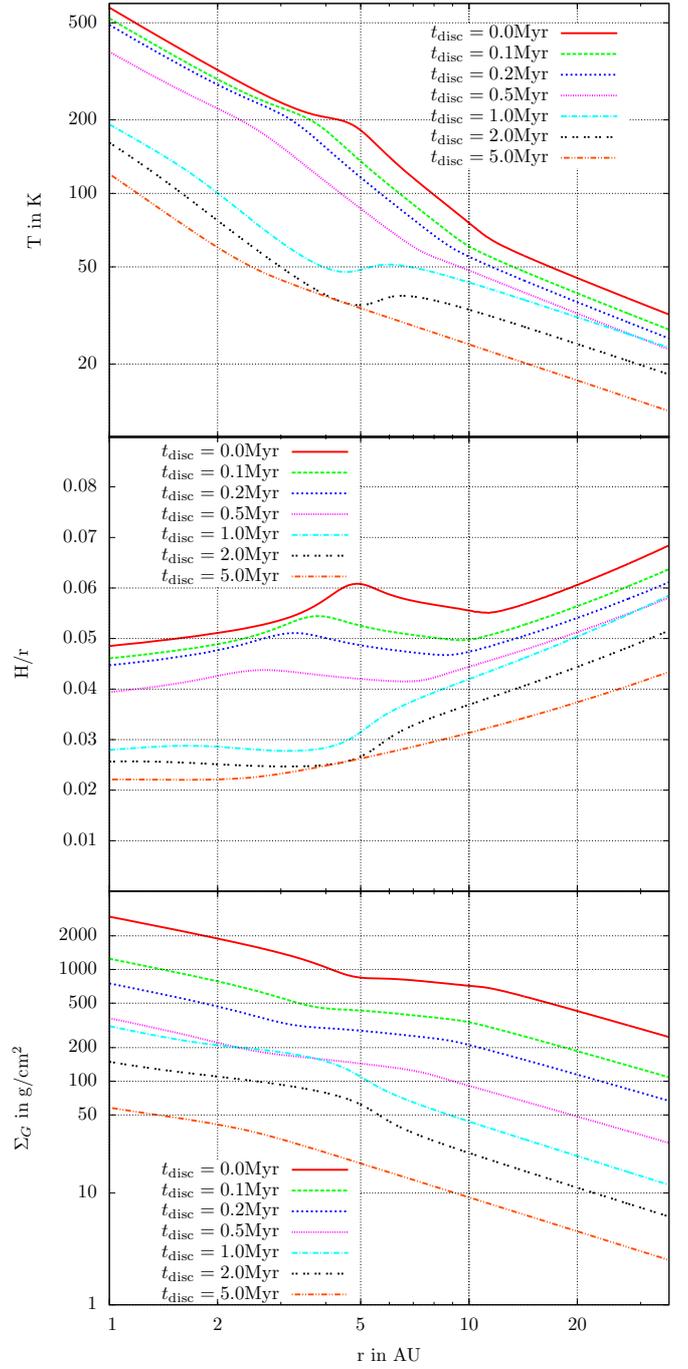}
 \caption{Mid-plane temperature (top), $H/r$ (middle), and surface density $\Sigma_G$ (bottom) for discs at different time evolutions. The values for $T$, $H/r$, and $\Sigma_G$ have been calculated from eq.~\ref{eq:Thigh} to eq.~\ref{eq:Tlow} with a metallicity of $0.5\%$. As time evolves, the surface density reduces and with it the inner region that is dominated by viscous heating, so that the disc is dominated by stellar irradiation at the late stages of the evolution, after a few Myr. In this late stage, the aspect ratio is nearly constant in the inner parts of the disc and $T$ and $\Sigma_G$ follow simple power laws.
   \label{fig:Mdottime}
   }
\end{figure}

To link these fits with time evolution, we relate back to eq.~\ref{eq:harttime}, which links time with a corresponding $\dot{M}$ value. For each $\dot{M}$ value, we can calculate the disc structure through eq.~\ref{eq:Thigh} to eq.~\ref{eq:Tlow} for discs with $Z \leq 0.005$ and through eq.~\ref{eq:Thighhm} to eq.~\ref{eq:Tlowhm} for discs with $Z>0.5$. We can therefore calculate what the disc structure looks like at different times. This is presented in Fig.~\ref{fig:Mdottime}, where temperature (top), $H/r$ (middle), and surface density (bottom) of discs with $Z=0.005$ and different ages are displayed.

The age of the disc, $t_{\rm disc}=0$ is not defined in eq.~\ref{eq:harttime}. Here we take the age $t_{\rm disc}=0$ to be $100$ kyr after the formation of the disc, because during the first stages, the disc is still undergoing massive changes in its structure \citep{2011ARA&A..49...67W}. We therefore recommend using this definition of $t_{\rm disc}=0$ age because it avoids discontinuities in the disc evolution. The age of the star, presented in table~\ref{tab:Starsize}, follows a different time scale, meaning age $t_{\rm disc}=0$ for the disc corresponds to a stellar age of $t_\star=100$ kyr. This means we add $100$ kyr in eq.~\ref{eq:harttime} to define the new $t_{\rm disc}=0$ age for the disc in the time of the star. Eq.~\ref{eq:harttime} then looks (without the measurement errors):
\begin{equation}
\label{eq:harttimenew}
 \log \left( \frac{\dot{M}}{M_\odot /\text{yr}} \right) = -8.00 - 1.40  \log \left( \frac{t_{\rm disc}+10^5\text{yr}}{10^6 \text{yr}} \right) \ .
\end{equation}
This equation describes the age of the disc $t_{\rm disc}$ in our set-up (indicated in Fig.~\ref{fig:Mdottime}), while eq.~\ref{eq:harttime} describes the age of the star $t_\star$ as stated in table~\ref{tab:Starsize}.

\bibliographystyle{aa}
\bibliography{Stellar}

\begin{thebibliography}{69}
\expandafter\ifx\csname natexlab\endcsname\relax\def\natexlab#1{#1}\fi

\bibitem[{{Alexander} {et~al.}(2014){Alexander}, {Pascucci}, {Andrews},
  {Armitage}, \& {Cieza}}]{2013arXiv1311.1819A}
{Alexander}, R., {Pascucci}, I., {Andrews}, S., {Armitage}, P., \& {Cieza}, L.
  2014, in {Protostars and Planets VI}, 2013arXiv1311.1819A

\bibitem[{{Alexander} \& {Pascucci}(2012)}]{2012MNRAS.422L..82A}
{Alexander}, R.~D. \& {Pascucci}, I. 2012, MNRAS, 442, pp.82

\bibitem[{{Andrews} {et~al.}(2010){Andrews}, {Wilner}, {Hughes}, {Chunhua}, \&
  {Dullemond}}]{2010ApJ...723.1241A}
{Andrews}, S.~M., {Wilner}, W.~J., {Hughes}, A.~M., {Chunhua}, Q., \&
  {Dullemond}, C.~P. 2010, ApJ, 723, pp.1241

\bibitem[{{Bai} \& {Stone}(2010{\natexlab{a}})}]{2010ApJ...722.1437B}
{Bai}, X.~N. \& {Stone}, J.~M. 2010{\natexlab{a}}, ApJ, 722, pp. 1437

\bibitem[{{Bai} \& {Stone}(2010{\natexlab{b}})}]{2010ApJ...722L.220B}
{Bai}, X.~N. \& {Stone}, J.~M. 2010{\natexlab{b}}, ApJ, 722, L220

\bibitem[{{Balbus} \& {Hawley}(1998)}]{1998RvMP...70....1B}
{Balbus}, S.~A. \& {Hawley}, J.~F. 1998, Reviews of Modern Physics, 70, 1

\bibitem[{{Baraffe} {et~al.}(1998){Baraffe}, {Chabrier}, {Allard}, \&
  {Hauschildt}}]{1998A&A...337..403B}
{Baraffe}, I., {Chabrier}, G., {Allard}, F., \& {Hauschildt}, P.~H. 1998, A\&A,
  337, p.403

\bibitem[{{Baruteau} {et~al.}(2014){Baruteau}, {Crida}, {Paardekooper},
  {Masset}, {Guilet}, {Bitsch}, Nelson, {Kley}, \&
  {Papaloizou}}]{2013arXiv1312.4293B}
{Baruteau}, C., {Crida}, A., {Paardekooper}, S.~J., {et~al.} 2014, in
  {Protostars and Planets VI}, arXiv:1312.4293

\bibitem[{{Baruteau} \& {Masset}(2008)}]{2008ApJ...672.1054B}
{Baruteau}, C. \& {Masset}, F. 2008, \apj, 672, 1054

\bibitem[{{Bell} \& {Lin}(1994)}]{1994ApJ...427..987B}
{Bell}, K.~R. \& {Lin}, D.~N.~C. 1994, ApJ, 427, 987

\bibitem[{Birnstiel {et~al.}(2012)Birnstiel, Klahr, \&
  Ercolano}]{2012A&A...539A.148B}
Birnstiel, T., Klahr, H., \& Ercolano, B. 2012, A\&A, 539, id.A148

\bibitem[{{Bitsch} {et~al.}(2013){Bitsch}, {Crida}, {Morbidelli}, {Kley}, \&
  {Dobbs-Dixon}}]{2013A&A...549A.124B}
{Bitsch}, B., {Crida}, A., {Morbidelli}, A., {Kley}, W., \& {Dobbs-Dixon}, I.
  2013, A\&A, 549, id.A124

\bibitem[{{Bitsch} \& {Kley}(2011)}]{2011A&A...536A..77B}
{Bitsch}, B. \& {Kley}, W. 2011, A\&A, 536, A77

\bibitem[{{Bitsch} {et~al.}(2014){Bitsch}, {Morbidelli}, {Lega}, \&
  {Crida}}]{2014A&A...564A.135B}
{Bitsch}, B., {Morbidelli}, A., {Lega}, E., \& {Crida}, A. 2014, A\&A, 564,
  id.A135

\bibitem[{Bitsch {et~al.}(2014)Bitsch, Morbidelli, Lega, Kretke, \&
  Crida}]{2014A&A...570A..75B}
Bitsch, B., Morbidelli, A., Lega, E., Kretke, K., \& Crida, A. 2014, A\&A, 570,
  id.A75

\bibitem[{{Brauer} {et~al.}(2008){Brauer}, {Henning}, \&
  {Dullemond}}]{2008A&A...487L...1B}
{Brauer}, F., {Henning}, T., \& {Dullemond}, C.~P. 2008, A\&A, 487, pp.L1

\bibitem[{Chiang \& Laughlin(2013)}]{2013MNRAS.431.3444C}
Chiang, E. \& Laughlin, G. 2013, MNRAS, 431, p.3444

\bibitem[{{Chiang} \& {Youdin}(2010)}]{2010AREPS..38..493C}
{Chiang}, E. \& {Youdin}, A. 2010, Annual Review of Earth and Planetary
  Sciences, 38, p.493

\bibitem[{{Chiang} \& {Goldreich}(1997)}]{1997ApJ...490..368C}
{Chiang}, E.~I. \& {Goldreich}, P. 1997, ApJ, 490, 368

\bibitem[{Cossou {et~al.}(2014)Cossou, Raymond, Hersant, \&
  Pierens}]{2014arXiv1407.6011C}
Cossou, C., Raymond, S.~N., Hersant, F., \& Pierens, A. 2014, astro-ph.EP,
  arXiv:1407.6011

\bibitem[{Fischer \& Valenti(2005)}]{2005ApJ...622.1102F}
Fischer, D.~A. \& Valenti, J. 2005, ApJ, 622, pp. 1102

\bibitem[{Fressin {et~al.}(2013)Fressin, Torres, Charbonneau, Bryson,
  Christiansen, Dressing, Jenkins, Walkowicz, \& Batalha}]{2013ApJ...766...81F}
Fressin, F., Torres, G., Charbonneau, D., {et~al.} 2013, ApJ, 766, id.81

\bibitem[{{Hartmann} {et~al.}(1998){Hartmann}, {Calvet}, {Gullbring}, \&
  {D'Alessio}}]{1998ApJ...495..385H}
{Hartmann}, L., {Calvet}, N., {Gullbring}, E., \& {D'Alessio}, P. 1998, ApJ,
  495, p.385

\bibitem[{{Hayashi}(1981)}]{1981PThPS..70...35H}
{Hayashi}, C. 1981, Progress of Theoretical Physics Supplement, 70, pp.35

\bibitem[{Ingleby {et~al.}(2014)Ingleby, Clavet, Hernandez, Briceno, Miller,
  Espaillat, \& McClure}]{arXiv:1406.0722}
Ingleby, L., Clavet, N., Hernandez, J., {et~al.} 2014, astro-ph.EP

\bibitem[{Johansen \& Lacerda(2010)}]{2010MNRAS.404..475J}
Johansen, A. \& Lacerda, P. 2010, MNRAS, 404, pp. 475

\bibitem[{{Johansen} {et~al.}(2007){Johansen}, {Oishi}, {Mac Low}, {Klahr},
  {Henning}, \& {Youdin}}]{2007Natur.448.1022J}
{Johansen}, A., {Oishi}, J.~S., {Mac Low}, M.~M., {et~al.} 2007, Nature, 448,
  pp. 1022

\bibitem[{{Johansen} \& {Youdin}(2007)}]{2007ApJ...662..627J}
{Johansen}, A. \& {Youdin}, A. 2007, ApJ, 662, pp. 627

\bibitem[{{Klahr} \& {Bodenheimer}(2003)}]{2003ApJ...582..869K}
{Klahr}, H.~H. \& {Bodenheimer}, P. 2003, \apj, 582, 869

\bibitem[{{Kley} {et~al.}(2009){Kley}, {Bitsch}, \&
  {Klahr}}]{2009A&A...506..971K}
{Kley}, W., {Bitsch}, B., \& {Klahr}, H. 2009, \aap, 506, 971

\bibitem[{{Kley} \& {Crida}(2008)}]{2008A&A...487L...9K}
{Kley}, W. \& {Crida}, A. 2008, \aap, 487, L9

\bibitem[{{Lambrechts} \& {Johansen}(2012)}]{2012A&A...544A..32L}
{Lambrechts}, M. \& {Johansen}, A. 2012, A\&A, 544, id.A32

\bibitem[{Lambrechts \& Johansen(2014)}]{2014arXiv1408.6094L}
Lambrechts, M. \& Johansen, A. 2014, ArXiv e-prints, arXiv:1408.6094

\bibitem[{Lambrechts {et~al.}(2014)Lambrechts, Johansen, \&
  Morbidelli}]{2014arXiv1408.6087L}
Lambrechts, M., Johansen, A., \& Morbidelli, A. 2014, astro-ph.EP,
  arXiv:1408.6087

\bibitem[{{Lega} {et~al.}(2014){Lega}, {Crida}, {Bitsch}, \&
  {Morbidelli}}]{2014MNRAS.440..683L}
{Lega}, E., {Crida}, A., {Bitsch}, B., \& {Morbidelli}, A. 2014, MNRAS, 440,
  p.683

\bibitem[{{Levermore} \& {Pomraning}(1981)}]{1981ApJ...248..321L}
{Levermore}, C.~D. \& {Pomraning}, G.~C. 1981, \apj, 248, 321

\bibitem[{{Levison} {et~al.}(2010){Levison}, {Thommes}, \&
  {Duncan}}]{2010AJ....139.1297L}
{Levison}, H.~F., {Thommes}, E., \& {Duncan}, M.~J. 2010, AJ, 139, pp.1297

\bibitem[{Lodders(2003)}]{2003ApJ...591.1220L}
Lodders, K. 2003, ApJ, 591, pp. 1220

\bibitem[{Martin \& Livio(2013)}]{2013MNRAS.434..633M}
Martin, R.~G. \& Livio, M. 2013, MNRAS, 434, p.633

\bibitem[{{Mihalas} \& {Weibel Mihalas}(1984)}]{1984frh..book.....M}
{Mihalas}, D. \& {Weibel Mihalas}, B. 1984, {Foundations of radiation
  hydrodynamics} (New York: Oxford University Press, 1984)

\bibitem[{Morbidelli {et~al.}(2009)Morbidelli, Bottke, Nesvorny, \&
  Levison}]{2009Icar..204..558M}
Morbidelli, A., Bottke, W.~F., Nesvorny, D., \& Levison, H.~F. 2009, Icarus,
  204, p.558

\bibitem[{{Morbidelli} \& {Nesvorny}(2012)}]{2012A&A...546A..18M}
{Morbidelli}, A. \& {Nesvorny}, D. 2012, A\&A, 546, id.A18

\bibitem[{{Nakagawa} {et~al.}(1986){Nakagawa}, {Sekiya}, \&
  {Hayashi}}]{1986Icar...67..375N}
{Nakagawa}, Y., {Sekiya}, M., \& {Hayashi}, C. 1986, Icarus, 67, p. 375

\bibitem[{{Nelson} {et~al.}(2013){Nelson}, {Gressel}, \&
  {Umurhan}}]{2013MNRAS.435.2610N}
{Nelson}, R.~P., {Gressel}, O., \& {Umurhan}, O.~M. 2013, MNRAS, 435, p.2610

\bibitem[{Ormel \& Klahr(2010)}]{2010A&A...520A..43O}
Ormel, C.~W. \& Klahr, H.~H. 2010, A\&A, 520, id.A43

\bibitem[{Ormel \& Kobayashi(2012)}]{2012ApJ...747..115O}
Ormel, C.~W. \& Kobayashi, H. 2012, ApJ, 747, id. 115

\bibitem[{{Paardekooper} {et~al.}(2010){Paardekooper}, {Baruteau}, {Crida}, \&
  {Kley}}]{2010MNRAS.401.1950P}
{Paardekooper}, S., {Baruteau}, C., {Crida}, A., \& {Kley}, W. 2010, \mnras,
  401, 1950+

\bibitem[{{Paardekooper} {et~al.}(2011){Paardekooper}, {Baruteau}, \&
  {Kley}}]{2011MNRAS.410..293P}
{Paardekooper}, S.~J., {Baruteau}, C., \& {Kley}, W. 2011, MNRAS, 410, 293

\bibitem[{{Paardekooper} \& {Mellema}(2006)}]{2006A&A...459L..17P}
{Paardekooper}, S.-J. \& {Mellema}, G. 2006, \aap, 459, L17

\bibitem[{{Pollack} {et~al.}(1996){Pollack}, {Hubickyj}, {Bodenheimer},
  {Lissauer}, {Podolak}, \& {Greenzweig}}]{1996Icar..124...62P}
{Pollack}, J.~B., {Hubickyj}, O., {Bodenheimer}, P., {et~al.} 1996, Icarus,
  124, 62

\bibitem[{{Rafikov}(2004)}]{2004astro-ph0406469}
{Rafikov}, R.~R. 2004, {\tt astro-ph/0406469}, \apj, submitted to ApJL

\bibitem[{Raymond \& Cossou(2014)}]{2014MNRAS.440L..11R}
Raymond, S.~N. \& Cossou, C. 2014, MNRAS, 440, L11

\bibitem[{{Raymond} {et~al.}(2009){Raymond}, {O'Brien}, {Morbidelli}, \&
  {Kaib}}]{2009Icar..203..644R}
{Raymond}, S.~N., {O'Brien}, D.~P., {Morbidelli}, A., \& {Kaib}, N.~A. 2009,
  Icarus, 203, p. 644

\bibitem[{{Ros} \& {Johansen}(2013)}]{2013A&A...552A.137R}
{Ros}, K. \& {Johansen}, A. 2013, A\&A, 552, id.A137

\bibitem[{Santos {et~al.}(2004)Santos, Israelia, \&
  Mayor}]{2004A&A...415.1153S}
Santos, N.~C., Israelia, G., \& Mayor, M. 2004, A\&A, 415, p.1153

\bibitem[{{Shakura} \& {Sunyaev}(1973)}]{1973A&A....24..337S}
{Shakura}, N.~I. \& {Sunyaev}, R.~A. 1973, \aap, 24, 337

\bibitem[{Sirono(2011)}]{2011ApJ...733L..41S}
Sirono, S.~I. 2011, ApJL, 733, L41

\bibitem[{Stoll \& Kley(2014)}]{2014arXiv1409.8429S}
Stoll, M.~H.~R. \& Kley, W. 2014, astro-ph.EP, arXiv:1409.8429

\bibitem[{{Tanaka} {et~al.}(2002){Tanaka}, {Takeuchi}, \&
  {Ward}}]{2002ApJ...565.1257T}
{Tanaka}, H., {Takeuchi}, T., \& {Ward}, W.~R. 2002, \apj, 565, 1257

\bibitem[{{Turner} {et~al.}(2014){Turner}, {Fromang}, {Gammie}, {Klahr},
  {Lesur}, {Wardle}, \& {Bai}}]{Turner2014}
{Turner}, N., {Fromang}, S., {Gammie}, C., {et~al.} 2014, in {Protostars and
  Planets VI}, arXiv:1401.7306

\bibitem[{{Walsh} {et~al.}(2011){Walsh}, {Morbidelli}, {Raymond}, {O'Brien}, \&
  {Mandell}}]{2011Natur.475..206W}
{Walsh}, K.~J., {Morbidelli}, A., {Raymond}, S.~N., {O'Brien}, D.~P., \&
  {Mandell}, A.~M. 2011, Nature, 475, pp. 206

\bibitem[{{Ward}(1997)}]{1997Icar..126..261W}
{Ward}, W.~R. 1997, Icarus, 126, 261

\bibitem[{Weidenschilling(1977)}]{1977MNRAS.180...57W}
Weidenschilling, S.~J. 1977, MNRAS, 180, p.57

\bibitem[{{Weidenschilling}(1977)}]{1977Ap&SS..51..153W}
{Weidenschilling}, S.~J. 1977, \apss, 51, 153

\bibitem[{Williams \& Cieza(2011)}]{2011ARA&A..49...67W}
Williams, J.~P. \& Cieza, L. 2011, Annual Review of Astronomy and Astrophysics,
  490, pp.67

\bibitem[{Yang {et~al.}(2009)Yang, {Mac Low}, \& Menou}]{2009ApJ...707.1233Y}
Yang, C.-C., {Mac Low}, M.~M., \& Menou, K. 2009, ApJ, 707, pp.1233

\bibitem[{Yang {et~al.}(2012)Yang, {Mac Low}, \& Menou}]{2012ApJ...748...79Y}
Yang, C.-C., {Mac Low}, M.~M., \& Menou, K. 2012, ApJ, 748, id.79

\bibitem[{Youdin \& Goodman(2005)}]{2005ApJ...620..459Y}
Youdin, A. \& Goodman, J. 2005, ApJ, 620, pp.459

\bibitem[{{Zsom} {et~al.}(2010){Zsom}, {Ormel}, {G{\"u}ttler}, {Blum}, \&
  {Dullemond}}]{2010A&A...513A..57Z}
{Zsom}, A., {Ormel}, C.~W., {G{\"u}ttler}, C., {Blum}, J., \& {Dullemond},
  C.~P. 2010, A\&A, 513, id.A57

\end{thebibliography}
\end{document}